\documentclass[twocolumn, nofloatfix]{aastex701}
\journalinfo{Accepted for publication in ApJ}

\usepackage{bm}
\usepackage{gensymb}

\usepackage{amsmath}

\usepackage{float}

\graphicspath{{figs}}

\begin{document}

\title{Three-dimensional magnetohydrodynamic shocks and associated plasma heating in the solar atmosphere}

\correspondingauthor{A. Giannis}  

\author[orcid=0009-0005-9065-2583]{A. Giannis}
\affiliation{Laboratory of Astronomy, Department of Physics, University of Ioannina, GR-45110, Ioannina, Greece}
\email[show]{a.giannis@uoi.gr}  

\author[orcid=0000-0002-6926-8676]{V. Archontis} 
\affiliation{Laboratory of Astronomy, Department of Physics, University of Ioannina, GR-45110, Ioannina, Greece}
\email{archontis@uoi.gr}

\begin{abstract}

We investigate the three-dimensional (3D) formation and evolution of magnetohydrodynamic (MHD) shock waves and their role in plasma heating within a solar bipolar active region. Using 3D compressible, resistive MHD simulations, we model the emergence of magnetic flux from the convection zone into a magnetized corona. Our results reveal that initial plasmoid ejections launch hot plasma jets preceded by slow-mode shocks. Unlike established 2D models, these ejections originate from multiple sites along the emerging field's periphery, resulting in shocks with a complex, inverted-U 3D morphology. Beneath these fronts, the heated plasma forms flattened, sheet-like columns. Furthermore, we find that sustained gas pressure gradients, driven by dense plasma accumulation from plasmoids, power a cool, dense upflow adjacent to the main reconnection jet. During the main reconnection phase, the collision of ejected plasmoids with the ambient field triggers the formation of fast-mode termination shocks. By mapping viscous heating directly onto these irregular 3D shock surfaces, we demonstrate that energy dissipation is highly localized. This finding provides a physical link between shock-induced viscous heating and the observed temperature enhancements at the apex of post-flare loops.

\end{abstract}

\section{Introduction}\label{sec:intro} 

It is generally acknowledged that the temperature of the solar corona is sustained at millions of degrees Kelvin by the dissipation of energy, which is injected from the subphotospheric layers of the Sun into its atmosphere by means of the Poynting flux. This energy is subsequently thermalized mainly through two dissipative channels: Ohmic dissipation of electric currents and viscous dissipation of both shear and compressive flows. While the precise mechanisms that heat the solar corona have yet to be fully understood, magnetic reconnection is held to be an important mechanism. This fundamental process has been proposed to operate on both small scales in the context of nanoflares \citep{parker_1988}, and on large-scale current sheets, where its role is observationally and theoretically established \citep{shibata_2011}. To this end, numerous works have attributed energy dissipation in the corona to Ohmic heating \citep[e.g.][]{priest_1998,gudiksen_2005,rappazzo_2008,gontikakis_2008,bingert_2011,kanella_2017,zhuleku_2021}.

Even so, recent studies emphasize that viscous dissipation can actually dominate in the corona’s very high magnetic Prandtl number regime \citep[see e.g.][]{Kumar_2006, Hood_2011, Mishonov_2011, Rempel_2017, Warnecke_2020}. Truly, a number of observational and modeling studies have demonstrated that viscosity can account for a significant fraction of the heating in solar active regions \citep[e.g.][]{Mahayan_2001, Litvinenko_2005, Chen_2022, Santos_2022, Daniel_2023, Thaler_2023, Zafar_2023}.  

Transient, eruptive events arise self-consistently in three-dimensional (3D) simulations of emerging flux regions \citep[e.g.][]{archontis_2004,moreno-insertis_2008,hood_2012,moreno-insertis_2013,archontis_2019,zhuleku_2025}. The models used in the above studies include viscosity and report heating and the formation of jets. Yet, although shock waves inherently form in these 3D environments, no study so far has fully characterized their 3D geometry nor studied the resulting highly intermittent viscous dissipation taking place at the shock front. A recent study by \citet{noraz_2025} identifies shocks and quantifies their contribution in heating, but focuses on the energy budget of the chromosphere in the quiet-Sun. Another series of works by \citet{bareford_2013,bareford_2015,bareford_2016,reid_2020} included shock-capturing viscosity algorithms in their simulations and demonstrated how this mechanism resulted in considerable energy dissipation to the system; they were however limited to the scope of kink-unstable coronal loops and have not addressed the precise identification of shocks. Bridging this gap in the literature is important, since shock fronts act as sites for rapid, localized energy dissipation and particle acceleration in the corona \citep{Vanninathan_2015,Matsumoto_2016,Fan_2021,Hou_2022,Kouloumvakos_2022,Kouloumvakos_2023,Manon_2024}.

Magnetic reconnection taking place in fully 3D environments is usually intermittent both in space and time \citep[][and references therein]{cheng_2023} and extends beyond two-dimensional (2D) reconnection models \citep[e.g.][]{sweet_1958,parker_1957,petschek_1964}, for which scaling relations for the reconnection rate can be sought. An important aspect of non-steady reconnection is the development of the resistive tearing instability \citep{furth_1963}, which is characterized by the formation of plasmoids (or magnetic islands) when the Lundquist number exceeds a critical value \citep{loureiro_2007}, which typically holds for the solar corona \citep{bhattacharjee_2009}. Notably, the formation of plasmoids in the context of active regions has been established, both in 2D simulations \citep[e.g.][]{yokoyama_1996,daniel_2025} -- where also their relation to coronal jets has been investigated -- as well as 3D ones \citep[e.g.][]{archontis_2006,moreno-insertis_2013}. These structures are known to be highly unstable; as per \citet{archontis_2006}, any imbalance in the reconnection rate along the current sheet gives rise to an imbalance in the Lorentz force, which in turn leads to the eventual ejection of the plasmoid via the ``melon-seed'' mechanism \citep[originally proposed by][see also \citealp{pneuman_1983,cargill_1984}]{schlueter_1957}. On this basis, \citet{shibata_2001} proposed the so-called plasmoid-induced reconnection model, which states that the ejection of a plasmoid from a 2D current sheet drives a strong inflow of plasma into the sheet, which results in fast magnetic reconnection. Since the number of plasmoids has been found to scale with the Lundquist number \citep{loureiro_2007}, the tearing instability naturally generates many plasmoids in coronal current sheets, and their ejection makes the energy release highly intermittent. Nonetheless, despite the expected strong velocity gradients during such impulsive reconnection events, the link between plasmoid-driven shocks and viscous heating has not been assessed in previous studies on 3D emerging flux.

The present work addresses the formation of shock waves in a 3D simulated solar active region. An attempt to classify the mode of the MHD shocks is made, along with a mapping of the related viscous dissipation. The paper is organized as follows. In Section \ref{sec:setup}, we describe the governing equations, our numerical setup, and the initial conditions of the simulation. A description of how the code handles shocks is also included. Section \ref{sec:events} focuses on identifying shocks and linking them with viscous dissipation in the solar corona during selected transient events from our 3D run. Section \ref{sec:concl} contains a summary and a discussion of our findings, as well as several future prospects.

\section{Numerical scheme}\label{sec:setup}

\subsection{Equations}

The numerical simulation presented in this paper is performed using the Lare3d code \citep{arber_2001}. The code solves the compressible, time-dependent, resistive MHD equations in Lagrangian form, which read in dimensionless units

\begin{align}
   \frac{D\rho}{Dt} &= -\rho (\bm{\nabla} \cdot \bm{v}), \label{eq:continuity}\\
   \rho \frac{D\bm{v}}{Dt} &= (\bm{\nabla} \times \bm{B}) \times \bm{B} - \bm{\nabla} P + \rho \bm{g} + \bm{F}_v, \label{eq:momentum} \\
   \frac{D\bm{B}}{Dt} &= (\bm{B} \cdot \bm{\nabla}) \bm{v} - \bm{B} (\bm{\nabla} \cdot \bm{v}) - \bm{\nabla} \times (\eta \bm{\nabla} \times \bm{B}), \label{eq:induction}\\
   \rho \frac{D\varepsilon}{Dt} &= -P (\bm{\nabla} \cdot \bm{v}) + Q_J + Q_v, \label{eq:energy}
\end{align}

\noindent where $D/Dt\equiv\partial/\partial t + (\bm{v}\cdot\bm{\nabla})$ is the convective (or Lagrangian) derivative, $\rho$ the mass density, $\bm{v}$ the plasma velocity, $\bm{B}$ the magnetic field, $P$ the isotropic thermal pressure, $\bm{g}$ the gravitational acceleration, $\bm{F}_v$ the force per unit volume stemming from the shock-capturing viscosity formulation (see Section \ref{sec:shock-capt}), $\eta$ the plasma resistivity, $\varepsilon = P/\rho(\gamma - 1)$ the specific internal energy, with $\gamma=5/3$ being the ratio of specific heats, $Q_J=\eta J^2$ is the Ohmic heating term, with $\bm{J}$ being the electric current density, and $Q_v$ the volumetric viscous heating rate. For the normalization, the following photospheric reference values have been used: $L_0 = 0.18 \ \text{Mm}$, $B_0 = 300 \ \text{G}$ and $\rho_0 = 1.67\times 10^{-4}\ \text{kg}\ \text{m}^{-3}$. One then may calculate the reference Alfv\'{e}n speed: $v_{A_0}=B_0/\sqrt{\mu_0 \rho_0} = 2.1 \ \text{km}\ \text{s}^{-1}$ (with $\mu_0 = 4\pi\times 10^{-7}\ \text{H}\ \text{m}^{-1}$ the permeability of the vacuum), as well as the reference Alfv\'{e}n time: $t_{A_0}=L_0/v_{A_0} = 86.9\ \text{s}$, which serve as the units of velocity and time for the normalization, respectively. The reference pressure is $P_0 = B_0^2/\mu_0 = 716\ \text{Pa}$, and the reference temperature is defined as $T_0 = \bar{m} v_{A_0}^2 / k_B = 649\ \text{K}$, or equivalently, as $T_0 = P_0 M_{\mathrm{mol}}/(\rho_0 R_{\mathrm{gas}})$, where $\bar{m} = \mu m_u$ is the average mass per particle in the gas mixture, with $m_u$ being the atomic mass constant, $k_B$ the Boltzmann constant, $M_{\mathrm{mol}}$ the molar mass, $M_{\mathrm{mol}} = \mu m_u N_A$, $N_A$ Avogadro's constant, and $R_{\mathrm{gas}} = N_A k_B$ the molar gas constant. We assume a constant mean molecular weight of $\mu = 1.25$. The resistivity is explicit ($\eta=0.01$ in dimensionless units or $4.75\ \Omega\cdot\text{m}$ in SI units) and uniform in the computational domain.

Lare3d utilizes a \textbf{La}grangian \textbf{re}map scheme in order to solve eqs. \eqref{eq:continuity}--\eqref{eq:energy}. Essentially, it splits each timestep into a Lagrangian step followed by a remap step onto the original Eulerian grid. The Lagrangian step uses a predictor-corrector scheme that is second-order accurate both in space and time. The physical quantities are then mapped back to the Eulerian grid, while the preservation of monotonicity is ensured by the implementation of van Leer gradient limiters \citep{vanLeer_1997}. Additionally, the variables are defined on a staggered grid which prevents numerical issues such as checkerboard patterns in the solution and at the same time facilitates the incorporation of conservation laws directly into the finite difference method.

\subsection{Shock-capturing}\label{sec:shock-capt}
It is desired for any viscosity formulation employed in magnetohydrodynamic (MHD) simulations to incorporate a shock-capturing mechanism \citep{Ryu_1995, Toth_2000,arber_2001,Mignone_2007}, thereby ensuring that viscous dissipation is connected to the irreversible conversion of kinetic energy at shocks into thermal energy. This is needed for an accurate resolution of steep discontinuities associated with wave steepening.

Numerical frameworks like Lare3d are ideal for shock capturing for mainly two reasons. Firstly, the Lagrangian step deforms the grid in such a way that it moves with the plasma. This allows steep velocity gradients, driven by Lorentz forces or gas pressure gradients, to shear and compress the local grid cells. These rapid variations in velocity lead to entropy production, which is accounted for through shock-capturing viscosity. The second reason is that gradient limiters help regularize the solution at shock fronts by damping spurious oscillations near discontinuities, which further ensures stability in the numerical scheme. 

Lare3d implements a shock-capturing viscosity formulation based on the works of \citet{kuropatenko_1967} and \citet{wilkins_1980} that introduces an additional force-per-volume term in the momentum equation \eqref{eq:momentum} and a heating term in the internal energy equation \eqref{eq:energy}. To achieve this, the numerical scheme computes a viscous term $q_{\text{Kur}}$, which has pressure dimensions and is inspired by the pressure jump condition across a shock:

\begin{equation}\label{eq:qkur}
\begin{split}
   q_{\text{Kur}} & = \rho \left[\nu_2 \frac{\gamma + 1}{4}|\Delta\mathbf{v}| \ + \right. \\ 
   & + \left. \sqrt{\nu_2^2 \left(\frac{\gamma + 1}{4}\right)^2 (\Delta\mathbf{v})^2 + \nu_1^2 c_s^2}\right] |\Delta\mathbf{v}|,
\end{split}
\end{equation}

\noindent where $\rho$ is the local mass density, $c_s=\sqrt{\gamma P/\rho}$ is the sound speed, $\Delta \mathbf{v}$ is the velocity jump across a grid cell, and $\nu_1$ and $\nu_2$ are dimensionless coefficients that control the viscosity. In the present simulation, we adopt $\nu_1=0.1$ and $\nu_2=0.2$. This formulation effectively combines a linear viscosity term (proportional to $\nu_1$), which is active even in regions of smooth flow to prevent numerical oscillations, and a quadratic viscosity term (proportional to $\nu_2$), which dominates at strong shocks to capture non-linear discontinuities. The required viscous force $\bm{F}_v$ and the corresponding volumetric viscous heating rate $Q_v$ are computed directly from $q_{\text{Kur}}$. Specifically, the formulation is based on the edge-viscosity scheme of \citet{caramana_1998}, where the viscous forces are calculated along cell edges using $q_{\text{Kur}}$ and applied to the grid nodes, and the volumetric heating rate $Q_v$ is defined as the work done by these forces (see Appendix A of \citet{reid_2020} for the detailed numerical implementation on a 3D staggered grid). This approach models the physical dissipation of kinetic energy into heat, ensuring a stable solution and robust shock capturing. 

\subsection{Simulation setup}

\begin{figure}[htbp]
   \centering
   \includegraphics[width=\linewidth]{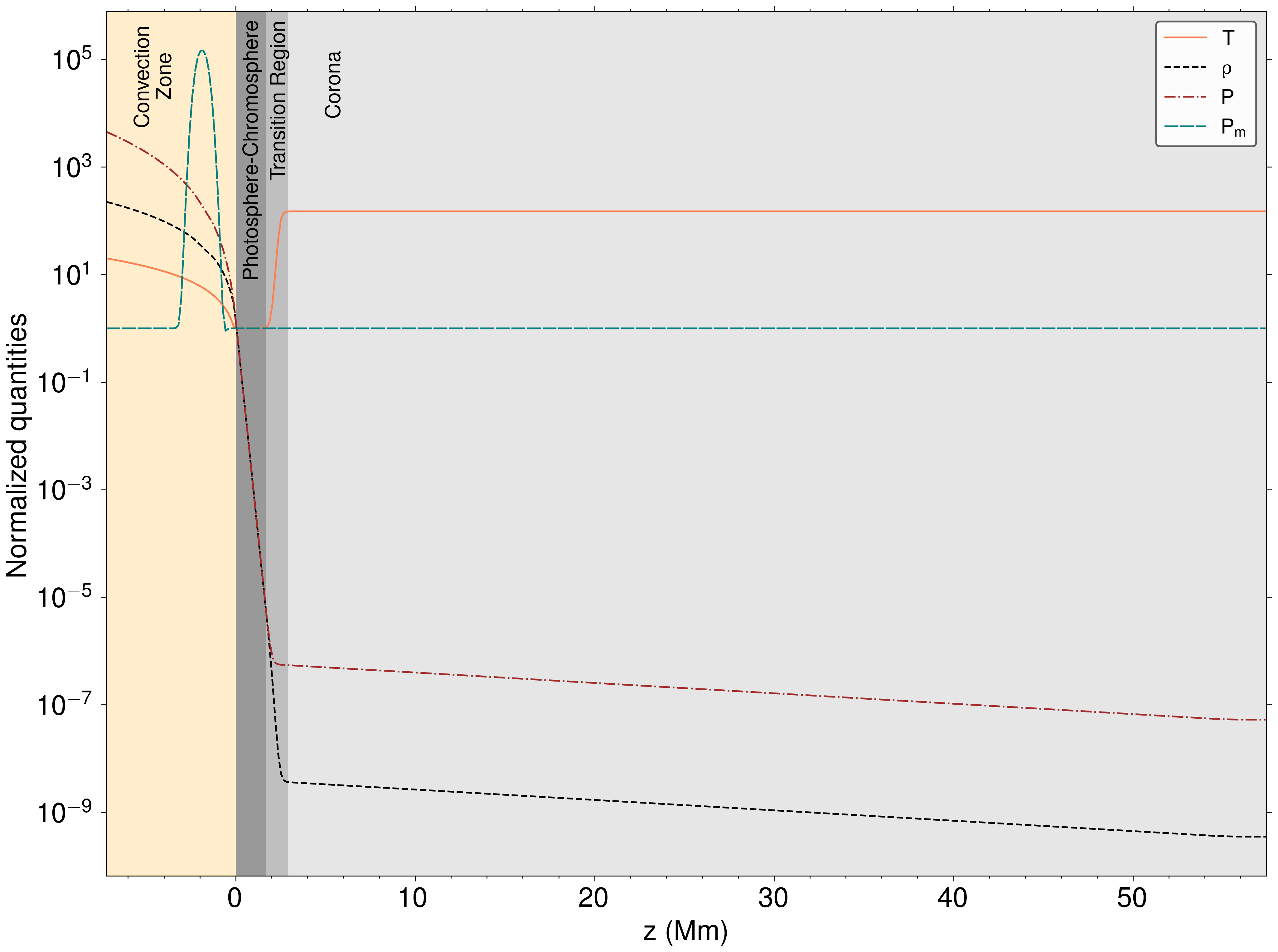}
   \caption{Initial vertical profiles of the normalized temperature (orange), mass density (black), gas pressure (brown), and magnetic pressure (teal) at $x = y = 0$, depicting the plane-parallel atmospheric stratification (where each layer is represented by a vertical band). The peak of the magnetic pressure corresponds to the subphotospheric magnetic flux tube at $z = -1.8\ \text{Mm}$.}
   \label{fig:init_stratif}
\end{figure}

The computational domain is a cartesian box of size 64.8 $\times$ 64.8 $\times$ 64.8 Mm in the $x, y, z$ directions, in a 420 $\times$ 420 $\times$ 420 uniform grid. The adiabatically stratified convection zone extends at $-7.2\ \text{Mm}\leq z < 0 \ \text{Mm}$, the isothermal photosphere-chromosphere at $0\ \text{Mm}\leq z < 1.8 \ \text{Mm}$, the transition region at $1.8\ \text{Mm}\leq z < 3.2 \ \text{Mm}$ and the isothermal corona at $3.2\ \text{Mm}\leq z < 57.6 \ \text{Mm}$. We employ open boundary conditions in the $x$-direction and at the top of the numerical box to allow waves and expanding plasma to leave the computational domain with minimal reflections. Periodic boundary conditions are used in the $y$-direction (along the axis of the magnetic flux tube) to prevent boundary effects, while the lower $z$-boundary is closed. The initial vertical profiles of temperature, mass density, gas pressure and magnetic pressure, all normalized by their photospheric ($z = 0$) values, are shown in Fig.~\ref{fig:init_stratif}.

A cylindrical, twisted magnetic flux tube in pressure equilibrium is placed 1.8 Mm below the photosphere, with its axis being oriented along the y-direction. For the magnetic field components of the flux tube we adopt the prescription of \citet{fan_2001}, namely

\begin{align}
B_r &= 0, \\
B_\phi &= \alpha r B_y,\\
B_y &= B_0 \exp (-r^2/R_0^2),
\end{align}

\noindent where $r$ is the radial distance from the tube axis, $\phi$ is the azimuthal direction, $B_0 = 3900\ \text{G}$ is the magnetic field strength, $R_0 = 0.45 \ \text{Mm}$ is the tube radius, and $\alpha=0.4\ (0.0023\ \text{km}^{-1})$ is a measure of the twist per unit of length. The flux tube is perturbed from its equilibrium state by imposing a density deficit along its axis, per \citet{fan_2001, archontis_2004}:

\begin{equation}\label{eq:density_deficit}
\Delta\rho = \frac{P_t(r)}{P(z)}\rho(z)\exp(-y^2/\lambda^2),
\end{equation}

\noindent where $P$ is the external pressure of the unperturbed stratification, $P_t$ is the total pressure within the flux tube and $\lambda=5 \ (0.9\ \text{Mm})$ is the length scale of the buoyant part of the flux tube. The atmosphere is magnetized with a uniform ambient magnetic field of strength $B_a = 10\ \text{G}$ which is oriented with $\theta=11 \degree$ and $\phi=180 \degree$, where $\theta$ and $\phi$ are the familiar spherical coordinate angles.
We finally highlight that the numerical model assumes a uniform mean molecular weight of $\mu = 1.25$ throughout the entire computational domain for simplicity (hence, the ionization state is not modeled dynamically). We note that there exist recent works with a similar numerical setup that include a more sophisticated, partially ionized plasma treatment \citep[e.g.][]{chouliaras_2023,moraitis_2024,chouliaras_2025}.


\section{Results}\label{sec:events}

The rise of the magnetic flux tube towards the magnetized atmosphere is driven by the imposed density deficit at the central part of the tube axis. This part of the tube becomes buoyant and emerges through the convection zone with an $\Omega$-loop-like shape. After approximately 16 minutes, the tube reaches the photosphere. At that point, it briefly slows down and only resumes its ascent towards the corona when a secondary buoyancy instability is triggered \citep{acheson_1979}, which does not occur until around $t = 22 \ \text{min}$. All of these initial processes have been reported in detail in numerous works \citep[e.g.][]{archontis_2004,magara_2004,galsgaard_2007, moreno-insertis_2008,moreno-insertis_2013, chouliaras_2023}. As our focus lies on the events that occur after the flux tube interacts with the coronal magnetic field, we will not discuss the details of these initial phases.

\begin{figure}[htbp]
   \centering
   \includegraphics[width=\linewidth]{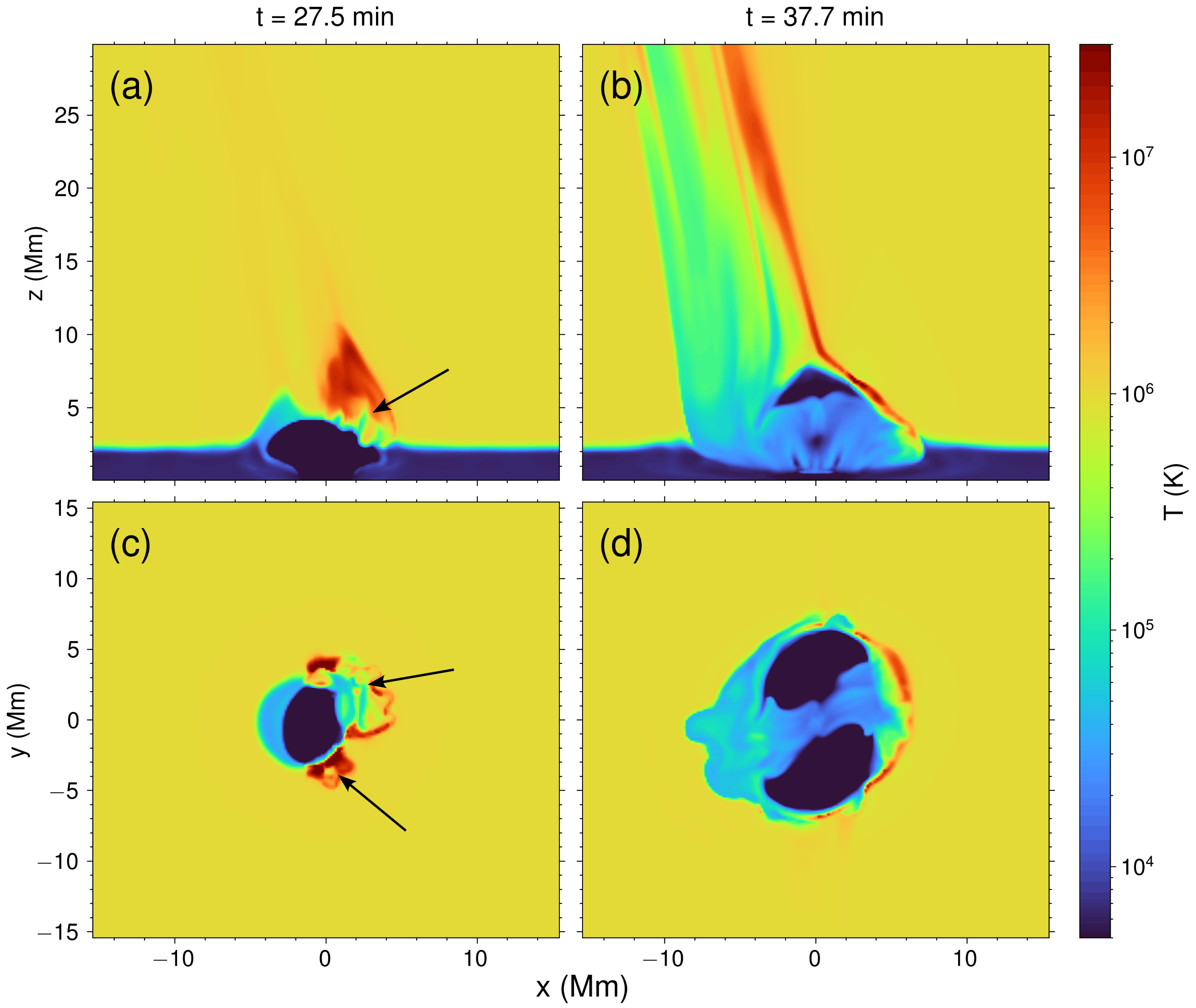}
   \caption{\textbf{Top row:} vertical temperature distribution at the $y=0$ mid-plane (normal to the magnetic flux tube axis). \textbf{Bottom row:} horizontal temperature distribution in the low corona at $z=3.6\,$Mm. The left column depicts the initial plasmoid ejection phase (with black arrows marking the sites where plasmoid ejections occur), while the right column depicts the main reconnection phase that leads to the formation of the ``standard'' jet. The colormap to the right shows the temperature in Kelvin.}
   \label{fig:overview}
\end{figure}

Figure \ref{fig:overview} shows the dynamical evolution of the simulation. Following the emergence of the magnetic flux tube in the corona, the first major event occurs at $t \sim 27 \ \text{min}$ (panels (a) and (c) of Figure \ref{fig:overview}) as plasmoids are ejected from the interface between the emerging and coronal magnetic field (ejection sites are indicated by black arrows). Afterwards, an inverse-Y-shaped coronal jet (henceforward referred to as a ``standard'' jet, following \citealt{moore_2010}) is launched at $33\ \text{min}\lesssim t \lesssim 43\ \text{min}$ as a direct result of magnetic reconnection between the two magnetic flux systems (panels (b) and (d) of Figure \ref{fig:overview}). After the ``standard'' jet, the system enters an advanced phase of its dynamical evolution, during which further jets are launched. These are characterized as ``blowout'' jets \citep{moore_2010}. In this scenario, a secondary, sheared or twisted magnetic flux system becomes unstable and erupts, ``blowing'' open the overlying field and driving both external interchange reconnection (which launches the jet) and internal reconnection (which forms a miniature flare arcade at the jet base). Hereafter, we limit our study to the evolution of the system until $t\sim 36\ \text{min}$. Our aim is to identify MHD shocks and study how energy is dissipated on their 3D structures as the system evolves during the aforementioned time period.

\subsection{Initial plasmoid ejection phase}\label{sec:init_plasmoids}

\begin{figure*}
\centering
\includegraphics[scale=0.86]{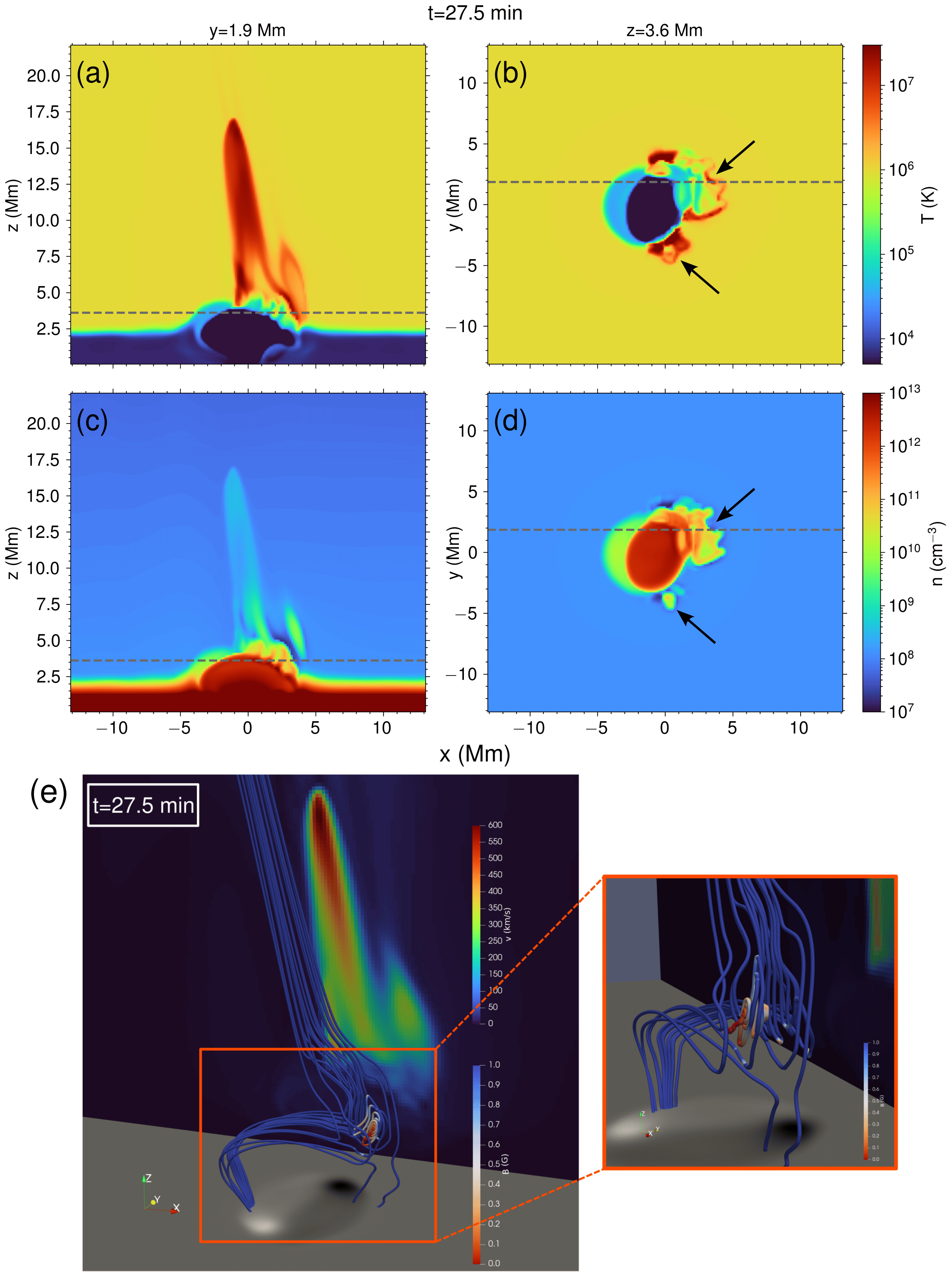}
\caption{2D distributions of the temperature (panels (a) and (b)) and number density (panels (c) and (d)) for the initial plasmoid ejection phase ($t=27.5\ \text{min}$). The left column depicts the vertical distribution of the two quantities on the plane $y=1.9\ \text{Mm}$, where the jet appears, while the right column depicts the horizontal distribution of the two quantities on the plane $z=3.6\ \text{Mm}$, which is located in the low corona. The black arrows in panels (b) and (d) mark plasmoid ejection sites. The gray dashed line in panels (a) - (d) indicates the location of the cross-section shown in the other column. Panel (e) provides a 3D illustration of the magnetic field line topology at $t=27.5\ \text{min}$. The field lines are colored based on the magnetic field magnitude, saturated at 1 Gauss to enhance the visibility of the plasmoid. The slice shows the vertical distribution of the velocity magnitude for $y=1.9\ \text{Mm}$, translated from its original position by $4\ \text{Mm}$ for clarity. The horizontal plane (grayscale) shows the distribution of the vertical magnetic field at the photosphere. The inset in panel (e) is a close-up of the plasmoid at the interface between emerging and ambient field.}
\label{fig:init_plasmoid_gen}
\end{figure*}

Figure \ref{fig:init_plasmoid_gen} illustrates the temperature (panels (a) and (b)) and a representative number density (panels (c) and (d), calculated as the mass density divided by the proton mass) distribution during the ejection of plasmoids from the interface between the emerging and the ambient field. Panel (a) shows that at around $z=3.6$ Mm (horizontal dashed line) and $0\ \text{Mm}\lesssim x \lesssim 4$ Mm, there is relatively cool material just above the expanding volume of the emerging field. This material is also dense (panel (c)). By tracing magnetic field lines around this area (panel (e)), we find that there are plasmoids with highly twisted field lines which are formed there and are ejected out of the interface and along the direction of the ambient field. The vertical slice in panel (e) shows plasma accelerating along the coronal field lines ahead of the ejecting plasmoids, reaching velocities up to approximately $600\ \text{km}\ \text{s}^{-1}$. As the plasmoids move upwards, they reconnect with the background magnetic field, and thus they lose their highly twisted nature. During their ejection, they bring dense material in the corona (panel (c)) and, as they reconnect, hot plasma is ejected upwards with them. Panels (b) and (d) show the temperature and density distribution at $z=3.6$ Mm. The visualization reveals that the ejection of hot material does not happen in a strictly 2D manner. It occurs all along the interface between the emerging and the coronal magnetic fields, which extends along the $y$-direction and mainly for positive $x$ (right-hand side of the emerging dome). For instance, in panels (b) and (d) we see at least two sites associated with plasmoid ejections, one at $-1\ \text{Mm}\lesssim y\lesssim 4$ Mm and one at $-5\ \text{Mm}\lesssim y\lesssim -2$ Mm, marked by black arrows. The formation and ejection of plasmoids start at around $t=27$ min and continue until $t=29$ min.

Since plasmoid ejections have been previously associated with the formation of shocks \citep{zenitani_2011, yang_2013, kumar_2023}, we proceed by investigating shock signatures during this event. The main diagnostics that we employ are: (i) the Lare3d viscous heating, which is essentially the volumetric heating rate $Q_v$ from eq. \eqref{eq:energy}, (ii) the plasma velocity field divergence: $\bm{\nabla}\cdot\bm{v}$ and (iii) the Mach number with respect to the local slow magnetosonic speed (i.e. slow-mode speed, \citealt{priest_2014}):

\begin{equation}\label{eq:slow-mach}
   M_s = \frac{v}{v_s},\ \text{where:} \ v_s^2 = \begin{cases} 
      v_A^2 & \text{for}\quad  c_s > v_A \\
      c_s^2 & \text{for}\quad  c_s < v_A
   \end{cases},
\end{equation}

\noindent since the velocity of the disturbance in question is well above that speed. In eq. \eqref{eq:slow-mach}, $v_s$ is the slow-mode speed, $v_{A}=B/\sqrt{\mu_0\rho}$ the Alfv\'en speed and $c_s$ the sound speed. We should note that for the calculation of $v_s$, we adopted the assumption of parallel (to the ambient field lines) propagation, which is reasonable for this particular event. As the coronal plasma is low-beta, this disturbance propagates as a sonic wave along the ambient field lines.

\begin{figure*}[t!]
\centering
\includegraphics[width=\textwidth]{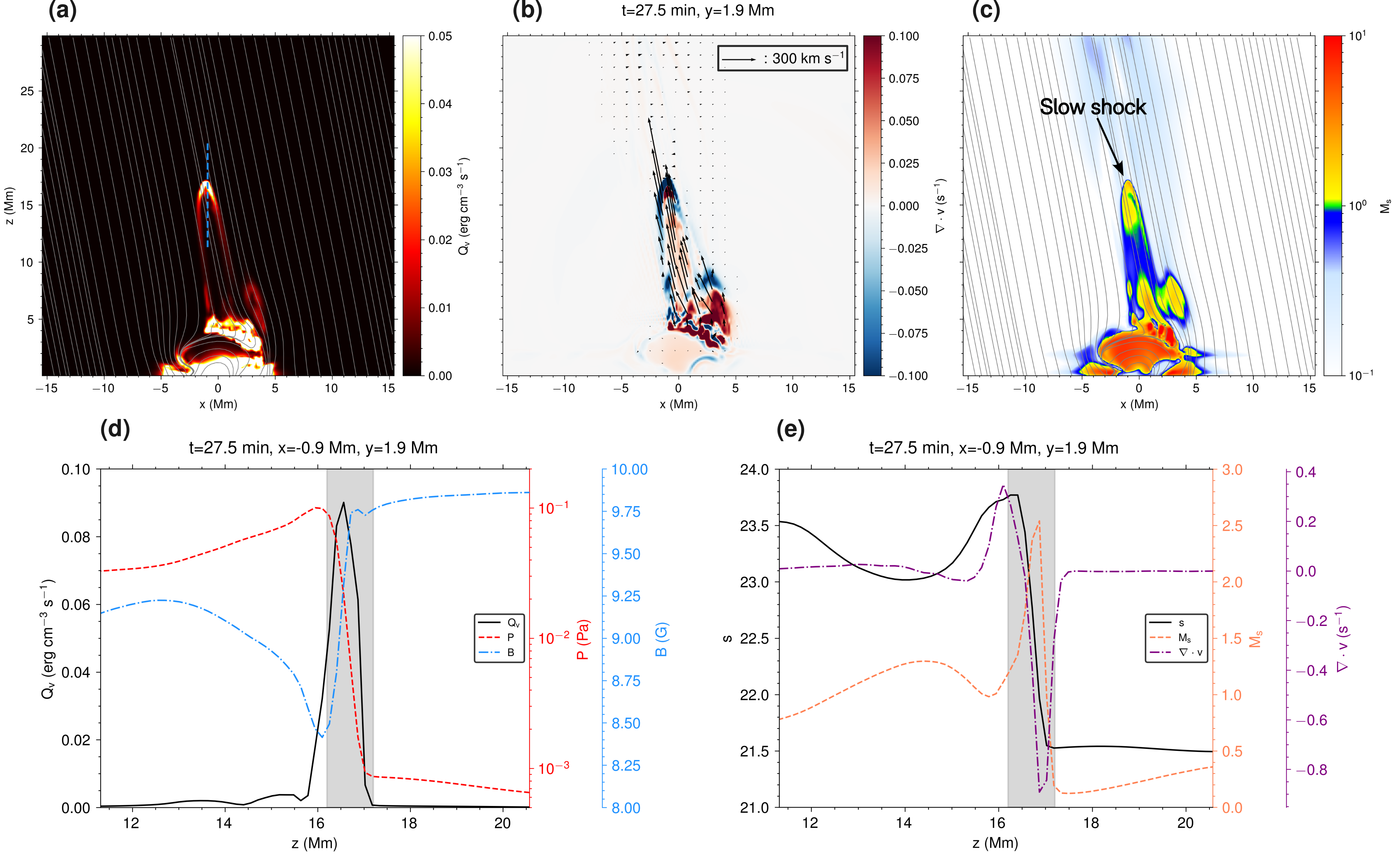}
\caption{MHD shock diagnostics for the initial plasmoid ejection phase ($t=27.5\ \text{min}$). \textbf{Top row:} 2D cuts at $y=1.9\ \text{Mm}$. (a) viscous heating $Q_v$, (b) divergence of the velocity field $\bm{\nabla}\cdot\bm{v}$, (c) slow-mode Mach number $M_s$. The color scale for the Mach number changes around $M_s = 1$ (green), signifying the transition to the super-magnetosonic regime (orange). Overlaid are the integral lines of the projection of the magnetic field vector (panels (a) and (c)) and the projection of the velocity vectors (panel (b)) on that plane. \textbf{Bottom row:} 1D cuts along the blue dashed line of panel (a). (d) Profiles of $Q_v$, thermal pressure $P$ and magnetic field magnitude $B$, (e) Profiles of the entropy proxy $s$, $M_s$ and $\bm{\nabla}\cdot\bm{v}$. The gray band denotes the approximate position of the shock.}
\label{fig:init_plasmoid_shock}
\end{figure*}

Figure \ref{fig:init_plasmoid_shock} (top row) is a visualization of $Q_v$, $\bm{\nabla}\cdot\bm{v}$ and $M_s$, at the same vertical slice as in Figure \ref{fig:init_plasmoid_gen} (a), (c).
In panel (a) of Figure \ref{fig:init_plasmoid_shock}, we find that viscous heating is very high at the front edge of the hot plasma jet. There, the plasma is experiencing strong compression (panel (b)) and it is also evident from panel (c) that the front propagates with a super-magnetosonic velocity ($v_{\text{shock}} > v_s$) relative to the unperturbed corona. These results indicate the presence of an MHD shock at the front of the jet propagating upwards into the solar atmosphere. To study this further, we analyze the profiles of various physical quantities along the blue dashed line shown in panel (a). In panel (d) we see that the peak in viscous heating correlates well with a sudden increase in the plasma thermal pressure by almost two orders of magnitude in the post-shock region. On the other hand, the magnetic field magnitude remains rather unchanged, showing a minor decrease of about $15\%$. These are characteristic signatures of a relatively strong slow-mode shock that propagates nearly parallel to the magnetic field lines, essentially as a sonic shock.

Furthermore, we study the distribution of the specific entropy of the plasma, $S$, of $M_s$ and of $\bm{\nabla}\cdot\bm{v}$. For the specific entropy of the plasma, we consider the relations $S = c_v \ln(P/\rho^\gamma) + \text{const.}$ and $\varepsilon = P/\rho(\gamma - 1)$, where $c_v$ is the specific heat at constant volume. Therefore, we can use a dimensionless entropy proxy that is equal to $s=\ln[(\gamma - 1)\varepsilon\rho^{1-\gamma}]$. Figure \ref{fig:init_plasmoid_shock}(e) shows an abrupt increase of the entropy proxy in the post-shock region.
The Mach number profile shows that it attains its maximum value, around 2.5, at the front of the shock. This correlates well with the highly negative values of the velocity divergence in this region. Overall, these results are consistent with previous studies, which demonstrated the upward launch of slow-mode shocks following the ejection of plasmoids \citep[e.g.][]{yang_2013}.

The above analysis provides a first indication for the formation of slow-mode shocks. Since the numerical simulation is 3D, we now present results regarding the actual 3D structure of the shock and the distribution of the heating in the shock region. To identify shock fronts in 3D, we employ the sonic-compression criterion \citep{wang_2020,finley_2022, noraz_2025} that reads:

\begin{equation}\label{eq:sonic-comp}
   \mathcal{S}\equiv \frac{-\bm{\nabla}\cdot\bm{v}}{c_s / (\epsilon\ dx)} > 1,
\end{equation}

\noindent where $dx$ is the distance between neighboring grid cells and $\epsilon$ is a dimensionless parameter that sets the characteristic thickness $\epsilon \ dx$ of the fronts classified as shocks. This condition serves as an empirical selection rule to locate shock fronts in the numerical model. Specifically, it compares the compression rate $(-\bm{\nabla}\cdot\bm{v})$ (which, if constant, represents the inverse timescale for a density growth by a factor $e$) to the inverse of the local sound-crossing time over a scale of $\epsilon\ dx$. By setting $\epsilon=4$, which is chosen to represent the characteristic scale of shock dissipation in the numerical scheme, we establish a conservative threshold that allows us to distinguish regions of sharp, non-linear compression (fronts thinner than $\lesssim 4$ grid cells) from high-amplitude but still linear acoustic waves. Tests on radiative MHD simulations \citep{noraz_2025} show that varying $\epsilon$ in the range $4-8$ changes the absolute fractions of the heating terms by at most tens of percent but leaves the spatial distribution and relative ordering of shock and current-sheet heating essentially unchanged. Our choice of $\epsilon=4$ is therefore sufficient for the present study, whose aim is to identify shocks and demonstrate how viscous dissipation is deposited on their surfaces, rather than to obtain an exact global heating budget.

\begin{figure*}[htb!]
\centering
\includegraphics[scale=1.09]{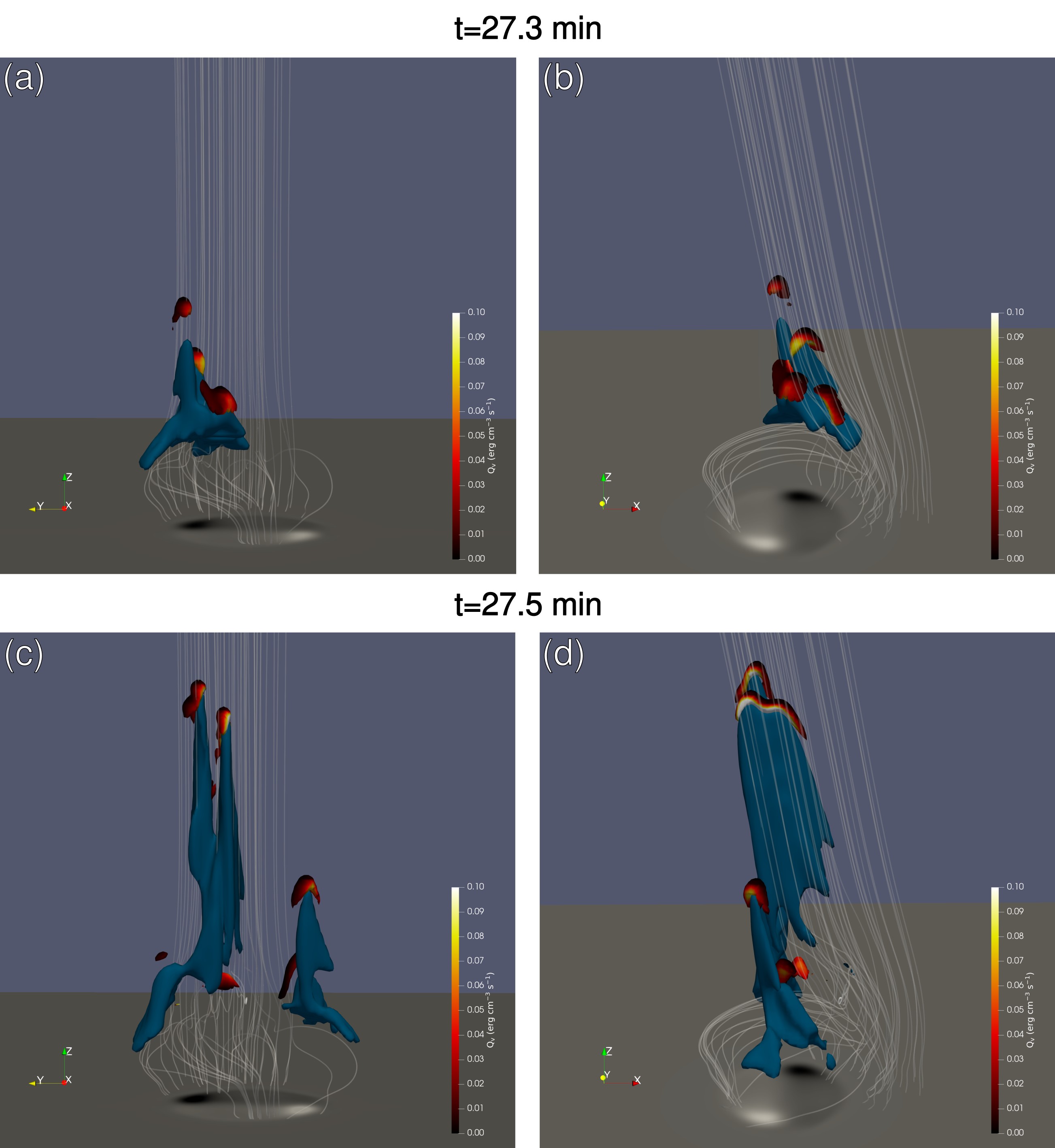}
\caption{3D morphology of slow-mode shocks and associated heating during the initial ejection phase. Isosurfaces of $\mathcal{S}=1.2$ (colored by the value of $Q_v$ in erg cm$^{-3}$ s$^{-1}$) are shown, along with high-temperature isosurfaces ($T=15\ \text{MK}$, in blue color). Several magnetic field lines are also traced in dim white color. The top row depicts the onset of the event ($t=27.3\ \text{min}$) from two different viewpoints (a) and (b), while the bottom row depicts a later stage of the event ($t=27.5\ \text{min}$) from the same two viewpoints (c) and (d). Note the characteristic inverted-U shape of the shock fronts and the localized nature of the viscous dissipation. The grayscale plane shows the photospheric vertical magnetic field. We also stress that the $y$-axis in panels (a) and (c) points to the left.}
\label{fig:init_plasmoid_shocks_3D}
\end{figure*}

Figure \ref{fig:init_plasmoid_shocks_3D} is a visualization of the 3D structure of the shocks together with areas of high temperature. The top panels are shown for $t=27.3\ \text{min}$, and the bottom panels for $t=27.5\ \text{min}$. More precisely, the more volume-filling structures in blue color represent isosurfaces with plasma temperature of $15\ \text{MK}$. On the other hand, the structures with lower spatial occupancy represent isosurfaces of $\mathcal{S}=1.2$ and they are color-coded according to the values of the viscous heating (colorbar). Magnetic field lines (in white color) have been traced so that they show the emerging loop and the ambient field. The horizontal plane shows the emerging bipolar region at the photosphere. Panels (a) and (c) show a y-z view of the structures' morphology and evolution. Panels (b) and (d) show the x-z view of the system.

At $t=27.3\ \text{min}$ (panels (a) and (b)), we show that very hot plasma is located at the interface between the emerging and the ambient magnetic field (panel (b)), where reconnection occurs between the field lines of the two magnetic flux systems. The hot plasma is found to be ejected along a laterally oriented Y-like junction and towards the negative-x direction. The ejection follows the reconnected field lines and, thus, it is moving upwards along the open ambient field and downwards along the outskirts of the emerging field. We are using the $\mathcal{S}=1.2$ isosurfaces to visualize where shocks occur in the 3D volume. They seem to appear in small, spatially localized regions distributed throughout the volume. We find that the shock isosurfaces ($\mathcal{S}=1.2$) appear near localized areas of the high-temperature ($T=15$~MK) isosurfaces. Also, the value of viscous heating is significantly enhanced within the shock isosurfaces relative to the surrounding medium, exhibiting localized maxima in some of the shocks.

Panels (c) and (d) show the same quantities visualized at a later time ($t=27.5\ \text{min}$). We find that the hot plasma continues to move along the reconnected field and it has now reached larger coronal heights, on the positive-y side of the emerging field. In comparison to the earlier stage (panels (a), (b)) we find some interesting features. The hot plasma associated with the jets is embedded within vertically oriented structures with a sheet-like three-dimensional morphology (panel (d)). These columns do not result from a single ejection event. For instance, panel (c) shows that there are two columns located in the positive-y half-space and one column located in the negative-y half-space, which is not apparent at earlier times (panel (a)). Because the jets occur intermittently and their onset does not occur simultaneously, the jet columns display a broad range of varying lengths, ranging from long, continuous structures while others appear shorter. Regarding the location of the shocks, we find that they are mainly at the front of the upwards propagating hot plasma outflows. This is actually what we have also shown in Figure \ref{fig:init_plasmoid_shock} (top panels), but in a 2D manner. The particular shock front which was identified using the 2D analysis, is basically located at the top of the vertical jet column located at the most positive $y$ (i.e. the leftmost column in Figure \ref{fig:init_plasmoid_shocks_3D}(c)) and has an inverted-U shape morphology with a relatively sharp apex. Next to it, there is another shock that develops small-scale corrugations during its upward propagation, although we do not assess whether these are associated with a corrugation instability \citep{stone_1995}. Similarly to previous times, viscous heating is locally enhanced at the shocks. Because the jets are preceded by shocks, viscous heating is continuously deposited along their propagation path. This process could provide a possible propagating source of plasma heating in the solar atmosphere.


\subsection{Formation of the cool column in the corona}\label{sec:upflow}

\begin{figure*}
\centering
\includegraphics[scale=0.61]{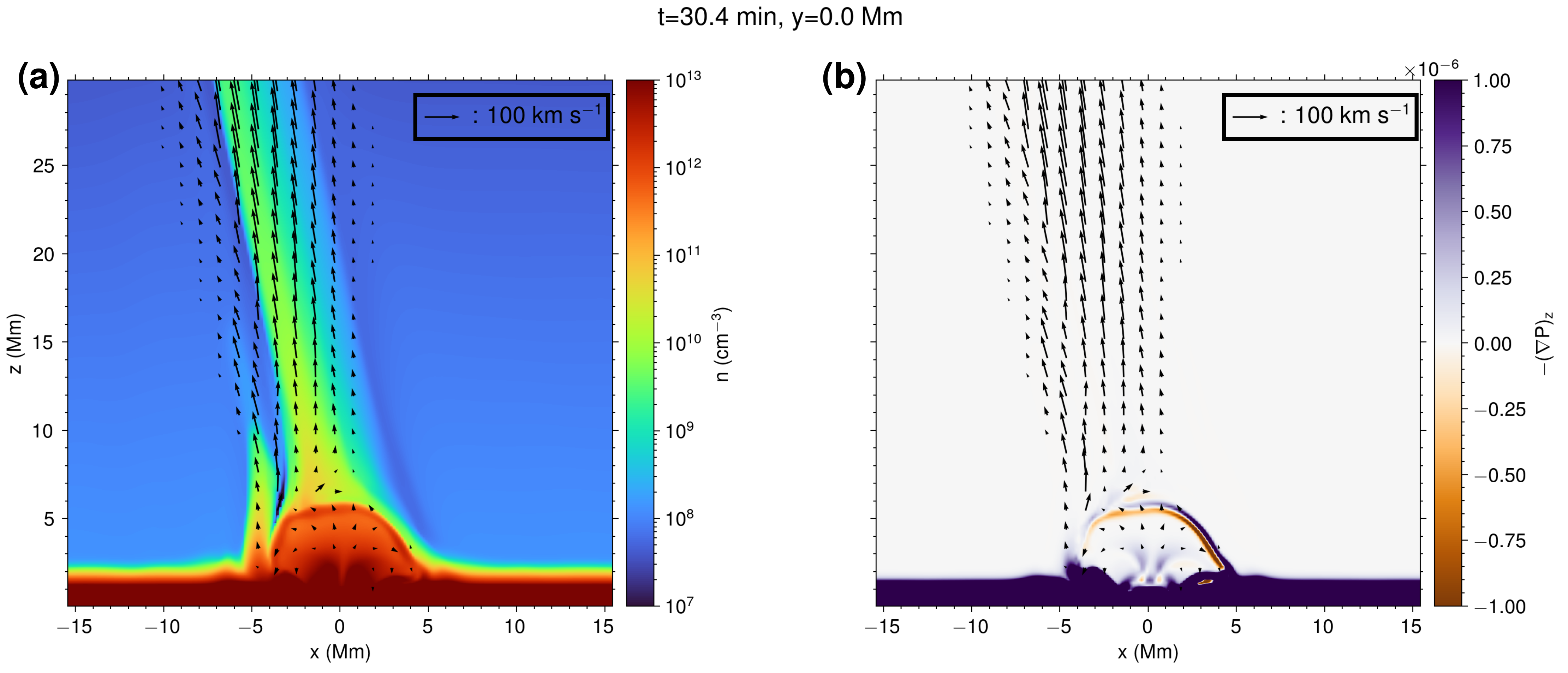}
\caption{Formation of the dense column in the corona. Vertical distribution of the number density (a) and of the vertical component of the gas pressure gradient (b) on the mid-plane $y=0$ for $t=30.4\ \text{min}$. The projection of the velocity field vectors is also overlaid on this plane in black arrows.}
\label{fig:plasmoid_upflow}
\end{figure*}

The formation and ejection of the plasmoids from the current sheet continue for a few more minutes before the emerging and the ambient fields come into direct contact and start reconnecting in a fast manner. Figure \ref{fig:plasmoid_upflow}(a) shows the density distribution at the vertical $xz$-mid-plane at $t=30.4\ \text{min}$. The projection of the velocity field vectors onto that plane is also overplotted. This selected snapshot is taken shortly after the launch of the hot plasmoid-driven jets described in Section \ref{sec:init_plasmoids}. Above the emerging field apex, there is a collimated outflow, which is dense and relatively cool. As the plasmoids get ejected, they start interacting (reconnecting) with the surrounding magnetic field and thus lose their high degree of twist. As a result, the dense material enclosed within them is released in the corona, and gets accelerated upwards with typical velocities on the order of a few tens of $\text{km}\ \text{s}^{-1}$. Since the plasmoids originate from the underlying atmospheric layers, specifically the transition region, this outflow has different physical properties (i.e. it has typical transition-region values of the temperature and density) compared to the hot jets discussed in Section \ref{sec:init_plasmoids}. We note that the plasmoids are not apparent in this snapshot because they were ejected shortly before $t=30\ \text{min}$.

On the left-hand side of the emerging volume, we find that there is another dense column at $x\sim -5\ \text{Mm}$ that extends up to $z\sim 10\ \text{Mm}$. The velocity field shows that the plasma in that column moves upwards with a vertical velocity $v_z\sim 70\ \text{km}\ \text{s}^{-1}$. This structure is reminiscent of the surges reported in previous studies, e.g. \citet{yokoyama_1996,moreno-insertis_2013,daniel_2016,nobrega_2017,nobrega_2018}. In our numerical experiment, we find that the heavy plasma of this column is transported there from the apex of the emerging field, and more precisely, via the continuous ejection of plasmoids. When the plasmoids are expelled from the current sheet, they are moving almost perpendicular (along the negative-$x$ direction) to the orientation of the ambient field. After the collision with the ambient field, the dense material of the plasmoids is redirected along, but also opposite to the direction of the ambient field. The latter motion deposits dense plasma in the column. We find that the accumulation of plasma in that area continues for as long as the formation and ejection of the plasmoids persist in the current sheet. By looking at the velocity field (black arrows) in both panels of Figure \ref{fig:plasmoid_upflow}, indeed, we see that next to the dense column there are downflows which are running all the way down to the base of the column. We will elaborate more on similar downflows, for a later time of the evolution in Figure \ref{fig:recon_plasmoid_shock}, where we will further show that at the collision site between this downflow and the heavy plasma at the low atmosphere a shock is formed. For now, we focus more on Figure \ref{fig:plasmoid_upflow}(b), which shows the $z$-component of the gas pressure gradient at the vertical mid-plane for $t=30.4\ \text{min}$. We find that the dense plasma deposition together with the highly stratified nature of the atmosphere leads to an area of high pressure gradient force at $-4.5\lesssim x\lesssim -3.5\ \text{Mm}$ and $1.5\lesssim z\lesssim 2.5\ \text{Mm}$, which is located at the base of the dense column and exactly below the upflow of the embedded plasma. This upflow seems to be sustained rather than be impulsive. We conjecture that the driving mechanism of the cool and dense plasma upflow is the pressure gradient force, which is triggered by the ejection of the plasmoids as we have discussed above.

\subsection{Main reconnection phase}\label{sec:plasmoids_recon}

\begin{figure*}[htb!]
\centering
\includegraphics[scale=1.27]{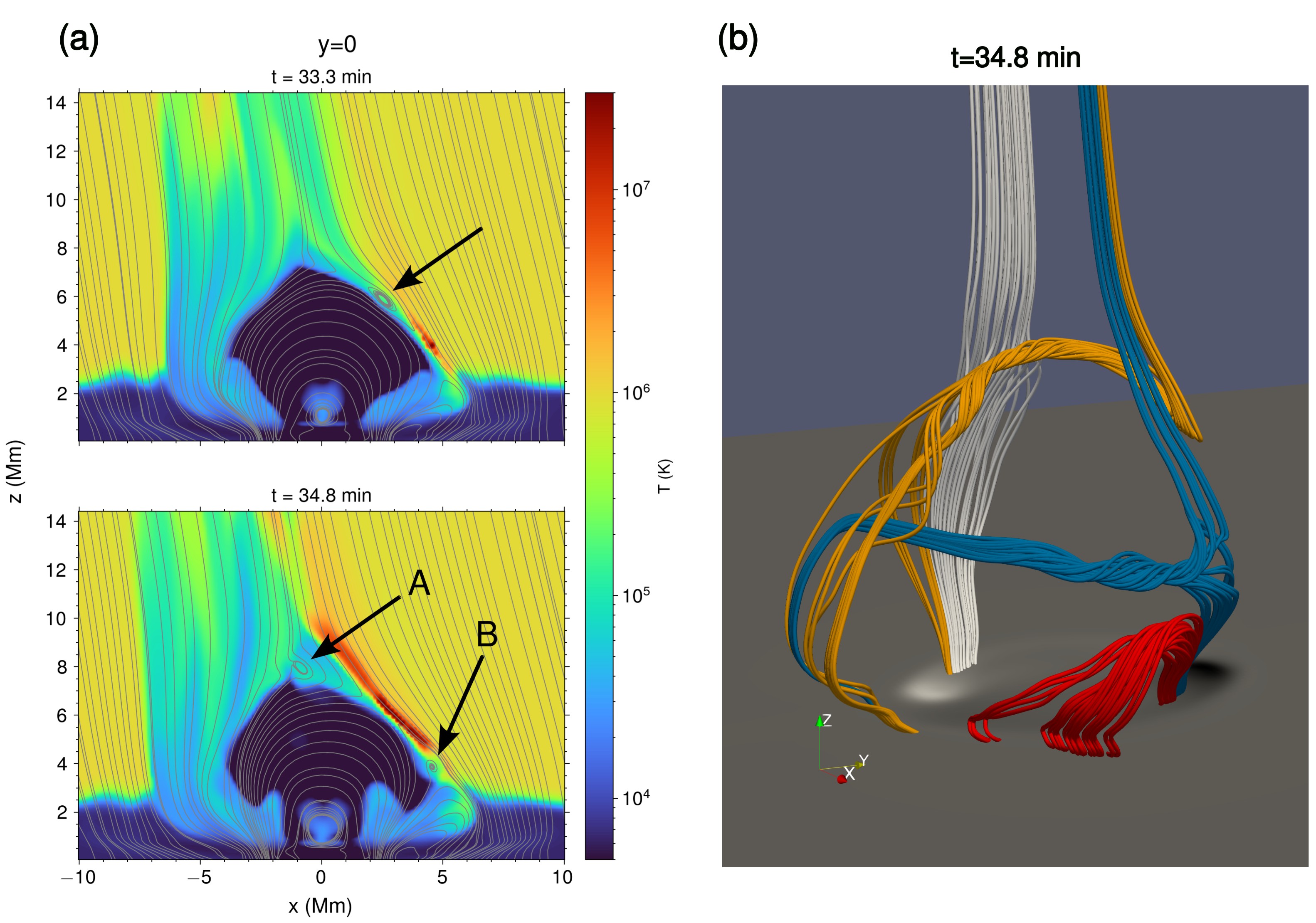}
\caption{(a) Vertical temperature slices on the mid-plane $y=0$, showing the onset of fast reconnection. \textbf{Top:} $t=33.3\,$min. \textbf{Bottom:} $t=34.8\,$min. Overplotted are the integral lines of the projection of the magnetic field vector on that plane in gray. The black arrows mark ejecting plasmoids. (b) 3D illustration of the plasmoid ejection for $t=34.8\,$min. The yellow and blue field lines are traced through the two plasmoids A and B, respectively, which are shown in the bottom panel of (a). The red field lines depict the closed loops, while the white ones represent the open reconnected field lines. The grayscale plane shows the photospheric vertical magnetic field.}
\label{fig:recon_plasmoid_generic}
\end{figure*}

Soon after the propagation of the initial slow-mode shocks, a ``standard'' jet is launched to the corona due to reconnection of the emerging field with the ambient coronal field along the interface of the two systems (panels (b) and (d) of Figure \ref{fig:overview}). This 3D reconnection produces a magnetic topology comprising both open field lines and closed field lines rooted in the photosphere, which may account for post-flare loops \citep{shibata_2011}. The topology of such 3D configurations has thoroughly been described in previous studies \citep[e.g.][]{moreno-insertis_2008,moreno-insertis_2013,archontis_2013,chen_2025}. Figure \ref{fig:recon_plasmoid_generic} shows the onset of the process which leads to the formation of the ``standard'' jet. Energy dissipation becomes first evident around $33\ \text{min}$ as a localized increase in plasma temperature at $(x, z)\approx(4.5,4.0)$ Mm (top panel of Figure \ref{fig:recon_plasmoid_generic}(a)), which intensifies further around $35\ \text{min}$ (bottom panel of Figure \ref{fig:recon_plasmoid_generic}(a)).
This is the region where a strong current sheet is built at the interface between the emerging and the ambient field. During this phase, plasmoids are formed anew within the current sheet (marked by the black arrows in Figure \ref{fig:recon_plasmoid_generic}(a)) and eventually, they are ejected out of it from both edges of the interface. By $34.8$ min, we detect the ejection of two plasmoids (bottom panel of Figure \ref{fig:recon_plasmoid_generic}(a)), at $(x_1, z_1) \approx (-1.0, 8.0)$ Mm and $(x_2, z_2) \approx (4.5, 4.0)$ Mm (hereafter referred to as plasmoids A and B, respectively). During the dynamical evolution of the plasmoids, reconnection between the emerging and ambient magnetic field lines starts to occur at the current sheet and becomes more efficient when the plasmoids leave the current sheet, and allow a more direct contact between the interacting magnetic fields. Reconnection leads to plasma heating at the interface up to a few tens of $\text{MK}$ and to the fast ejection of hot plasma upwards (along the open reconnected magnetic field lines) and downwards (towards the closed loops at low coronal heights).

Figure \ref{fig:recon_plasmoid_generic}(b) provides a 3D visualization of the magnetic field lines at $t=34.8$ min. Here, we show four sets of field lines. The yellow field lines have been traced from the region within and at close proximity to plasmoid A. Similarly we have traced the field lines in blue color from plasmoid B. The field lines in white color have been traced from the left-hand side to plasmoid A and the red field lines from the right-hand side and below plasmoid B. When the emerging and ambient fields come into contact and reconnect, they create two reconnected sets of field lines like the ones shown in white and red color in Figure \ref{fig:recon_plasmoid_generic}(b). The 3D field line topology is more complicated when we take into account the motion of the plasmoids. During their ejection, the plasmoids interact with the ambient field and with other systems of previously reconnected field lines. For instance, plasmoid A interacts with the open ambient field lines on its left-hand side, after its ejection from the current sheet. In a similar manner, plasmoid B interacts with the ambient field and also with the closed loops of reconnected field lines at around $z=3$ Mm during its ejection. This process triggers magnetic reconnection at multiple sites and also one field line may undergo numerous episodes of reconnection during the evolution of the system. This could explain partially the complex connectivity of the field lines that pass through the two plasmoids shown in Figure \ref{fig:recon_plasmoid_generic}(b). One common feature in both sets of field lines (yellow and blue) is the highly twisted nature of the plasmoids. Due to the aforementioned complex interaction between the magnetic flux systems, we find that the yellow field lines (plasmoid A) establish connections with: (a) the footpoints of the open reconnected (white) field lines, (b) the open ambient field lines (yellow) creating a sharp V-shaped topology which reveals the high tension of the reconnected field lines therein; the sharp V shape is present in the traced 3D field lines and is not an effect of viewing perspective, and (c) the edges of the emerging flux system at around $(x, y) = (0.4, -5.0)$ Mm on the photospheric surface. The field lines through plasmoid B connect in a similar way to plasmoid A with the emerging and the ambient field, but also connect with the closed loops on the right-hand side of the emerging field (Figure \ref{fig:recon_plasmoid_generic}(a)).

Overall, the visualization of these field lines at that snapshot shows the complex connectivity between the magnetic flux systems in three dimensions. It also shows that in order to fully understand the dynamical evolution of the field line topology, a more detailed study is required, which is out of the scope of this study. However, we would like to explore more the connection between the plasmoids and the ``standard'' jet phase to the onset of shocks and the associated plasma heating which is discussed in the next paragraphs.

Previous studies in similar experiments \citep[e.g.][]{yokoyama_1996, Miyagoshi_2004,Takasao_2015, Takasao_2016, daniel_2016} have presented an analysis related to the location and onset of shocks, mainly in 2D and 2.5D numerical simulations. Here, we extend the analysis further by studying more conditions and requirements for the onset of the shocks and we also show a qualitative representation of the 3D structure and field line topology related to this subject.

\begin{figure*}[htb!]
\centering
\includegraphics[scale=0.82]{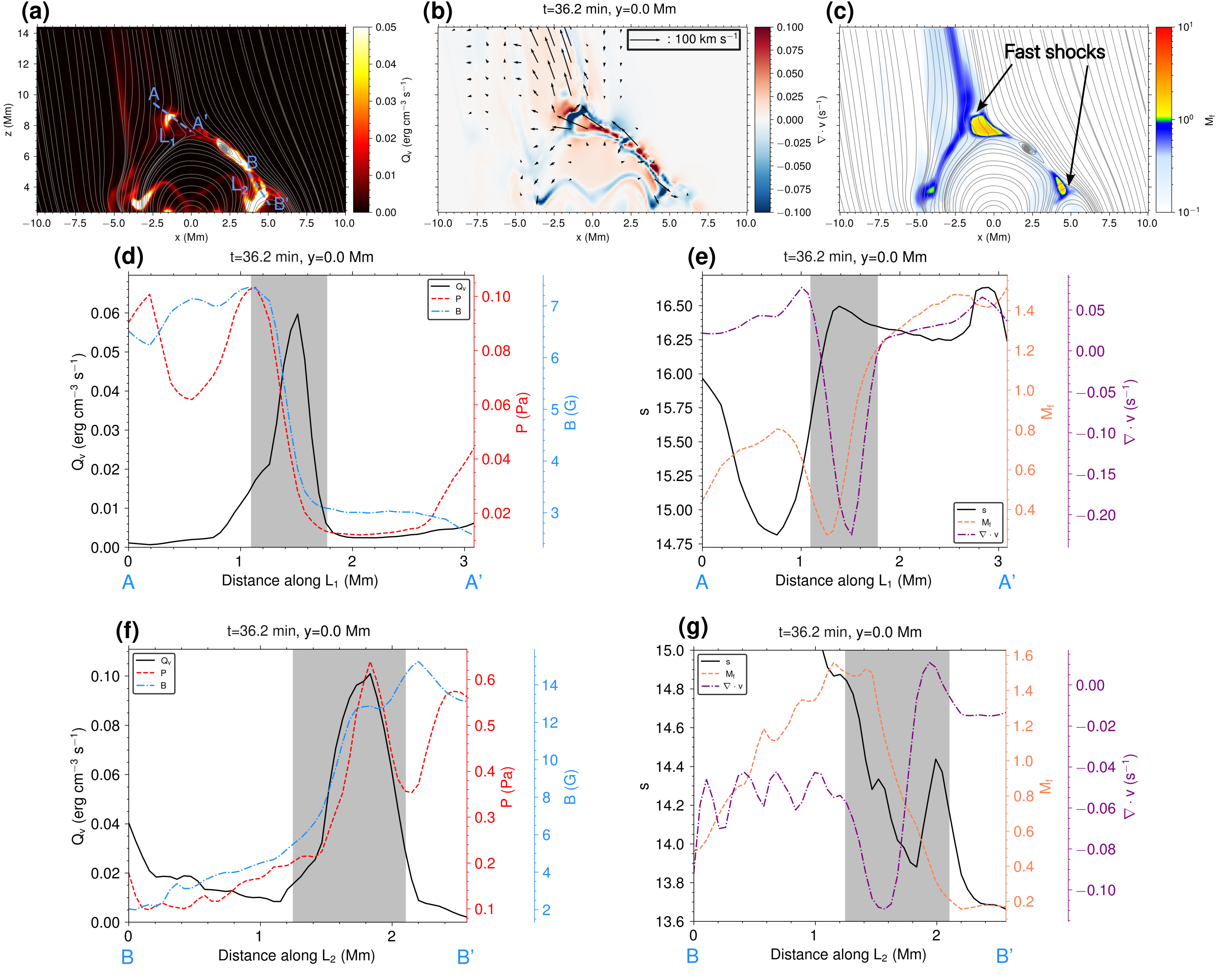}
\caption{MHD shock diagnostics for the main reconnection phase ($t=36.2\ \text{min}$, $y=0\ \text{Mm}$), in the same format and notation as Figure \ref{fig:init_plasmoid_shock}. \textbf{Top row:} 2D shock diagnostics, displaying (a) the viscous heating $Q_v$, (b) the velocity field divergence $\bm{\nabla}\cdot\bm{v}$, and (c) the fast-mode Mach number $M_f$. Panels (a) and (c) show integral lines of the projection of the magnetic field vector onto the $xz$ plane, while panel (b) shows projected velocity vectors. \textbf{Middle row:} 1D diagnostics along the line segment $L_1$, showing: (d) $Q_v$, thermal pressure $P$, and magnetic field magnitude $B$; (e) entropy proxy $s$, $M_f$, and $\bm{\nabla}\cdot\bm{v}$. \textbf{Bottom row:} 1D diagnostics along the line segment $L_2$, showing: (f) $Q_v$, $P$, and $B$; (g) $s$, $M_f$, and $\bm{\nabla}\cdot\bm{v}$. The gray bands denote the approximate positions of the two shocks.}
\label{fig:recon_plasmoid_shock}
\end{figure*}

We proceed by looking into the same shock diagnostics as with Section \ref{sec:init_plasmoids} with the aim of identifying shock structures in relation to the termination of the reconnection outflows and the ejection of plasmoids. For this purpose, we select two line segments on the mid-plane $y=0$ for $t=36.2\ \text{min}$ that are roughly parallel to the two outflows. These are visible in Figure \ref{fig:recon_plasmoid_shock}(a) as the two blue dashed line segments. The first segment $L_1$ is defined by the points $A=(-2.5, 9.5)\ \text{Mm}$ and $A'=(0.0, 7.6)\ \text{Mm}$, while the other segment $L_2$ is defined by $B=(3.7,5.0)\ \text{Mm}$ and $B'=(5.1,2.8)\ \text{Mm}$. Figure \ref{fig:recon_plasmoid_shock}(a) shows that viscous heating attains large values at four main locations: the one location is at the middle of the line AA' at $z\approx 8.5$ Mm. This specific area is located at the front of the ejected plasmoid A and in between the plasmoid and the open field lines at its left. The second location is at $5\lesssim z \lesssim 7$ Mm, again close to the edges of a new plasmoid which starts to be ejected out of the current sheet. The overplotted integral lines of the projection of the magnetic field vector onto the $xz$ plane show the core of the plasmoid. The third location of large viscous heating is found at heights between 2.5 and 4 Mm on the right hand side of the emerging field and at the top of the reconnected closed magnetic loops. The fourth location is on the left of the emerging field at $x\approx -3$ Mm and $2.5\lesssim z \lesssim 3.5$ Mm. Panel (b) of the same figure shows the compression of the plasma along with the projection of the velocity vector field, which shows the motion of the reconnection outflows and the plasmoids on that plane. At these four locations, we observe a general spatial association between regions of strong plasma compression and enhanced viscous heating. Panel (c) shows the Mach number for the same snapshot and plane. We calculate the fast-mode Mach number by assuming that the propagation is perpendicular ($\theta = 90^\circ$) with respect to the magnetic field lines, which is a reasonable approximation for the particular collision sites of the plasmoids with the surrounding magnetic field configuration. Under this assumption:

\begin{equation}\label{eq:fast-mach}
   M_f = \frac{v}{c_f} = \frac{v}{\sqrt{v_A^2 + c_s^2}},
\end{equation}

\noindent where $c_f$ is the fast magnetosonic speed \citep{priest_2014}. As shown in Figure \ref{fig:recon_plasmoid_shock}(c), the fast-mode Mach number is mildly super-magnetosonic ($M_f > 1$) within the outflows and decreases to sub-magnetosonic values as the plasma decelerates upon impacting the surrounding magnetic field configuration. This overall picture is less prominent for the second location, which happens mainly because the process of the plasmoid ejection has just started. 

Figure \ref{fig:recon_plasmoid_shock}(b) further shows that, as the ejected plasmoid A collides with the open reconnected field lines around the location of $L_1$, the flow in the post-shock region is split into two components, one moving upwards and one moving downwards. Although both of these two flows are sub-magnetosonic therein (visible as two blue lanes in Figure \ref{fig:recon_plasmoid_shock}(c)), the downflow (lower blue lane) follows the left-hand side of the emerging field periphery, and around $(x,z)=(-4, 3.5)$ Mm, the conditions are such that another shock is formed there (visible as a green bubble at that region). Below that region, the plasma gets compressed and heated by viscosity (panels (b) and (a) of Figure \ref{fig:recon_plasmoid_shock}). This region is actually associated with the upward motion of a column with cool and dense plasma, as discussed in Section \ref{sec:upflow}. 

\begin{figure*}
\centering
\includegraphics[scale=0.83]{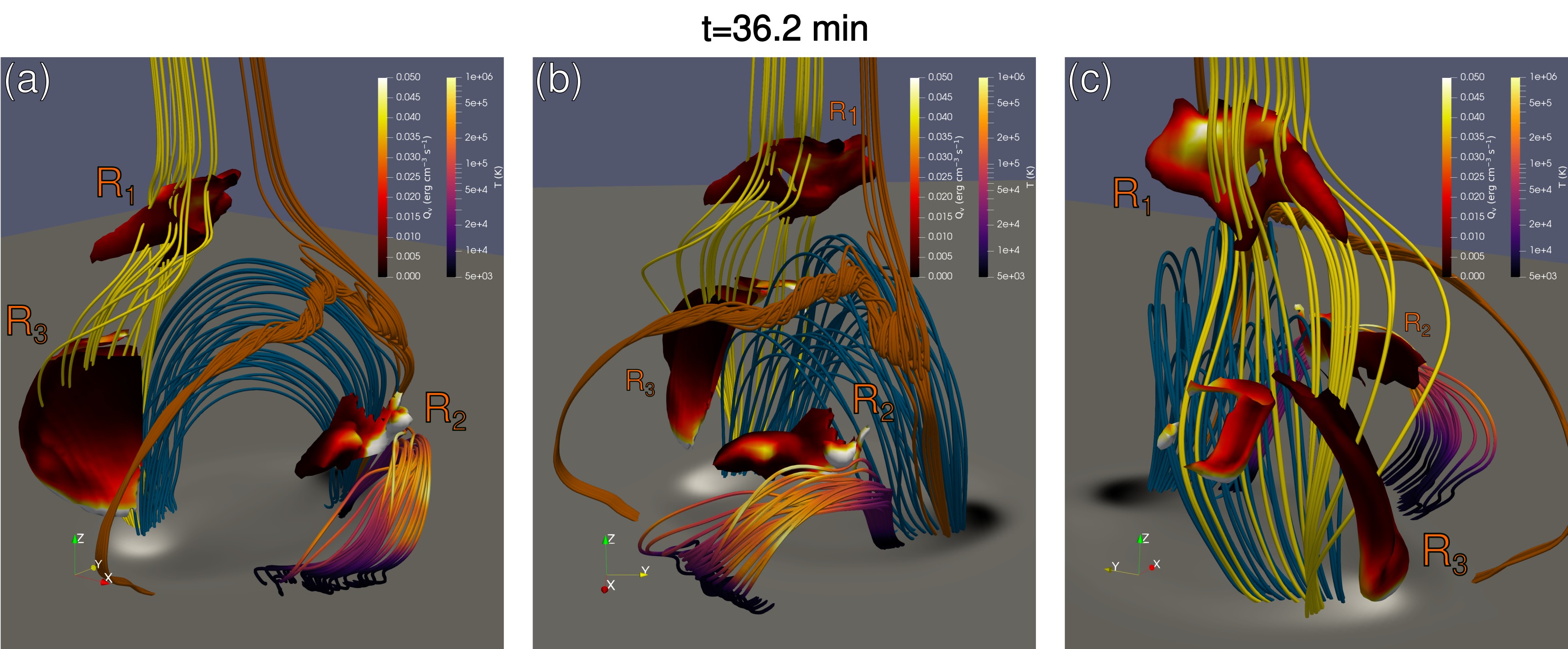}
\caption{3D morphology of the shocks and associated heating during the main reconnection phase. Isosurfaces of $\mathcal{S}=1.2$ (colored by the value of viscous heating in erg cm$^{-3}$ s$^{-1}$) for $t=36.2\ \text{min}$, drawn for three regions of interest; Region 1 ($R_1$), corresponding to the collision site with the open reconnected field lines, Region 2 ($R_2$), corresponding to the apex of the closed loops and Region 3 ($R_3$), corresponding to the left-hand side of the emerging field periphery. The magnetic field line topology is also illustrated, showing: the emerging field (in blue), the open reconnected field lines (in yellow), several field lines threading a plasmoid (brown), and the loops colored by the values of the temperature in Kelvin. Panels (a), (b), (c) represent different view points. The grayscale plane shows the photospheric vertical magnetic field.}
\label{fig:recon_plasmoid_shocks_3D}
\end{figure*}

In the middle and bottom rows of Figure \ref{fig:recon_plasmoid_shock}, we calculate one-dimensional profiles along the line segments $L_1$ and $L_2$, respectively. We have chosen these two locations as they are the most pronounced by that time of the evolution. In both cases, the magnetic field magnitude and the thermal pressure increase in the post-shock region (Figure \ref{fig:recon_plasmoid_shock}(d), (f)), signifying the presence of two fast-mode shocks. These transitions correlate spatially with peaks in viscous heating. Panels (e) and (g) of Figure \ref{fig:recon_plasmoid_shock} further show local minima of the velocity divergence and abrupt transitions of the fast-mode Mach number from super- to sub-magnetosonic velocities, as the plasma decelerates after crossing the shock. From these two panels, we find that in $L_1$, the plasma moves with a pre-shock velocity $v \sim 1.3c_f$, while in $L_2$ the downflow is ejected with $v \sim 1.5c_f$. For the shock along $L_1$ (Figure~\ref{fig:recon_plasmoid_shock}(e)), we observe both an entropy and temperature increase across the front. For the other case at $L_2$ (Figure \ref{fig:recon_plasmoid_shock}(g)), the entropy proxy also exhibits a jump in the apex of the loops, which is connected to the viscous heating that is deposited in them.

The analysis presented above focused more on the study of the shocks at $L_1$ and $L_2$ in Figure \ref{fig:recon_plasmoid_shock}(a). Here, we describe the topology of the magnetic field and the 3D structure of the shocks around these areas. Figure \ref{fig:recon_plasmoid_shocks_3D} is a visualization of the 3D shock isosurfaces with $\mathcal{S}=1.2$ (eq. \eqref{eq:sonic-comp}), colored by the values of the viscous heating. This snapshot is taken at the same time ($t=36.2\ \text{min}$) with the results presented in Figure \ref{fig:recon_plasmoid_shock}. The field lines have been traced from four different areas with the aim to show the topology of the magnetic field at that stage of the evolution. The blue field lines show the top part of the emerging field, which has not been reconnected yet. The yellow field lines have been traced just below the area at $L_1$ in Figure \ref{fig:recon_plasmoid_shock}(a) and they are the result of reconnection between the ejected plasmoid and the open ambient field, towards the negative $x$-direction. The brown field lines have been traced from within the current sheet at the interface between the emerging and ambient fields. The high degree of twist of these field lines reveals the existence of another plasmoid, which is about to be ejected out of the current sheet. The core of this plasmoid is visible in panels (a) and (c) of Figure \ref{fig:recon_plasmoid_shock} at $(x, z) =(2,6.5)\ \text{Mm}$. We see that the field lines from the plasmoid connect also the emerging with the ambient field since they have already experienced reconnection. The closed loops that are shown in various colors in panels (a) and (b) of Figure \ref{fig:recon_plasmoid_shocks_3D} are reconnected field lines formed after the interaction between ejected plasmoids and the ambient field on the right-hand side of the emerging field. These field lines have been colored according to the value of the temperature. At the top of the loops the temperature is high (a few MK) and the plasma is compressed locally by a fast shock (panel (c), Figure \ref{fig:recon_plasmoid_shock}). As we have mentioned before in this paper, these loops could be considered as post-flare loops.

The three panels in Figure \ref{fig:recon_plasmoid_shocks_3D} show also the 3D shock isosurfaces at three regions ($R_1$, $R_2$ and $R_3$). These regions correspond to the vicinity of the locations $L_1$ and $L_2$ in Figure \ref{fig:recon_plasmoid_shock}(a) and the location of the shock at the left of the emerging field (green bubble of Figure \ref{fig:recon_plasmoid_shock}(c)), respectively. They have been colored by the value of the viscous dissipation. All of the isosurfaces could be described as spatially extended, 3D plasma regions with irregular morphology. At the edges of $R_1$ and $R_2$ shock isosurfaces, we notice that corrugations develop over time but we have not studied their origin. Regarding the viscous heating, we find that its deposition is highly localized on the shock fronts. At $R_1$, viscous heating exhibits a localized increase at relatively large heights ($z \sim 8-9\ \text{Mm}$) within the reconnected (yellow) field lines, at $R_2$ it is concentrated at the top of the post-flare loops and at $R_3$ viscous heating is higher at low atmospheric heights (around $2\ \text{Mm}$).

\section{Conclusions and Discussion}\label{sec:concl}

The present study has addressed the formation of MHD shocks and the associated viscous dissipation in a 3D simulation of a solar emerging flux region, assuming the resistive MHD model. For the identification of the shocks, several diagnostics were used, namely the volumetric viscous heating rate, the velocity divergence, and the magnetosonic Mach number. These were further verified using 1D profiles to analyze the jumps of the physical quantities across the shock fronts. The 3D structure of the shocks was determined using the sonic-compression criterion (eq. \eqref{eq:sonic-comp}) and the viscous dissipation was subsequently mapped on the 3D shock isosurfaces. 

The main findings of this work are summarized below.

\begin{enumerate}
    \item The initial ejection of plasmoids from the current sheet occurs across multiple regions in 3D, and results in the launching of hot plasma in a jet-like manner in the corona. 
    \item Slow-mode shocks have been identified ahead of the hot plasma ejecta. These shocks have an inverted-U shape in 3D, and several of them exhibit corrugations. In the post-shock region, the heated plasma extends in a flattened, sheet-like shape.
    \item The formation of the cool and dense column adjacent to the hot ``standard'' jet has been attributed to strong gas pressure gradients, which form due to the accumulation of dense plasma in the lower atmosphere owing to the plasmoid ejection process.
    \item During the main reconnection phase, multiple plasmoids are intermittently ejected, leading to the formation of fast-mode termination shocks at collision sites with the surrounding magnetic field. These shocks exhibit an elongated, irregular 3D topology, moving beyond the idealized planar structures of 2D models.
    \item A major result of this study is the direct mapping of viscous heating onto 3D shock surfaces. We demonstrate that dissipation is concentrated in localized patches, rather than being uniformly distributed along the shock front. This localized heating provides a robust mechanism for the temperature increase observed at the apex of post-flare loops and suggests that shock-related viscosity is a primary channel for energy dissipation in the solar corona.
\end{enumerate}

The results presented in this study relate to earlier works in several noteworthy ways. For instance, although the upward launch of slow-mode shocks following plasmoid ejections has been established in previous 2D models \citep[e.g.][]{yang_2013}, our 3D simulation resolves the 3D structure of these shocks. Furthermore, our findings offer a different physical interpretation for the formation of the dense, cool plasma column that stands adjacent to the hot jet. While earlier studies have reported similar surge-like structures \citep[e.g.][]{yokoyama_1996,moreno-insertis_2013,daniel_2016,nobrega_2017,nobrega_2018}, our analysis attributes this sustained upflow to the plasmoid ejection process. In addition, the work by \citet{kengo_2023} has demonstrated that multiple shocks form in 3D above-the-loop-top regions during simulated solar flares. Our present work reveals how viscous heating is deposited onto the 3D surfaces of these termination shocks.

This study could be extended in several directions.
First of all, even though the present analysis focuses on the early dynamical evolution of the system, specifically the initial plasmoid ejections and the main reconnection phase up to $t\sim 36$ min, we intend to investigate the advanced eruptive phases in future work. Following the ``standard'' jet, the system's evolution is dominated by blowout jets, which are driven by the eruption of magnetic structures from the lower atmosphere. Applying the diagnostics presented in this work will reveal where shocks preferentially form during these eruptions. This will allow us to evaluate how viscous heating during these eruptions compares to the initial phases, alongside a comparative analysis of the successive eruptive events themselves. 
Furthermore, a statistical analysis of the viscous heating events identified in this work could shed light on the intermittent and turbulent nature of energy dissipation in the corona.
Lastly, the effects of partial ionization on the formation and heating of shocks have yet to be investigated in 3D simulations of emerging flux regions, and this could be addressed in a future study.


\begin{acknowledgments}
   The authors warmly thank the anonymous referee for their constructive review which substantially improved this manuscript. This research has been supported by the European Research Council through the Synergy Grant No. 810218 (“The Whole Sun”, ERC-2018-SyG).
\end{acknowledgments}


\bibliography{references}{}

\begin{thebibliography}{}
\expandafter\ifx\csname natexlab\endcsname\relax\def\natexlab#1{#1}\fi
\providecommand{\url}[1]{\href{#1}{#1}}
\providecommand{\dodoi}[1]{doi:~\href{http://doi.org/#1}{\nolinkurl{#1}}}
\providecommand{\doeprint}[1]{\href{http://ascl.net/#1}{\nolinkurl{http://ascl.net/#1}}}
\providecommand{\doarXiv}[1]{\href{https://arxiv.org/abs/#1}{\nolinkurl{https://arxiv.org/abs/#1}}}

\bibitem[{D.~J. {Acheson}(1979){Acheson}}]{acheson_1979}
{Acheson}, D.~J. 1979, \bibinfo{title}{{Instability by magnetic buoyancy.},} \solphys, 62, 23, \dodoi{10.1007/BF00150129}

\bibitem[{T.~D. {Arber} {et~al.}(2001){Arber}, {Longbottom}, {Gerrard}, \& {Milne}}]{arber_2001}
{Arber}, T.~D., {Longbottom}, A.~W., {Gerrard}, C.~L., \& {Milne}, A.~M. 2001, \bibinfo{title}{{A Staggered Grid, Lagrangian-Eulerian Remap Code for 3-D MHD Simulations},} Journal of Computational Physics, 171, 151, \dodoi{10.1006/jcph.2001.6780}

\bibitem[{V. {Archontis} {et~al.}(2006){Archontis}, {Galsgaard}, {Moreno-Insertis}, \& {Hood}}]{archontis_2006}
{Archontis}, V., {Galsgaard}, K., {Moreno-Insertis}, F., \& {Hood}, A.~W. 2006, \bibinfo{title}{{Three-dimensional Plasmoid Evolution in the Solar Atmosphere},} \apjl, 645, L161, \dodoi{10.1086/506203}

\bibitem[{V. {Archontis} \& A.~W. {Hood}(2013){Archontis} \& {Hood}}]{archontis_2013}
{Archontis}, V., \& {Hood}, A.~W. 2013, \bibinfo{title}{{A Numerical Model of Standard to Blowout Jets},} \apjl, 769, L21, \dodoi{10.1088/2041-8205/769/2/L21}

\bibitem[{V. {Archontis} {et~al.}(2004){Archontis}, {Moreno-Insertis}, {Galsgaard}, {Hood}, \& {O'Shea}}]{archontis_2004}
{Archontis}, V., {Moreno-Insertis}, F., {Galsgaard}, K., {Hood}, A., \& {O'Shea}, E. 2004, \bibinfo{title}{{Emergence of magnetic flux from the convection zone into the corona},} \aap, 426, 1047, \dodoi{10.1051/0004-6361:20035934}

\bibitem[{V. {Archontis} \& P. {Syntelis}(2019){Archontis} \& {Syntelis}}]{archontis_2019}
{Archontis}, V., \& {Syntelis}, P. 2019, \bibinfo{title}{{The emergence of magnetic flux and its role on the onset of solar dynamic events},} Philosophical Transactions of the Royal Society of London Series A, 377, 20180387, \dodoi{10.1098/rsta.2018.0387}

\bibitem[{M.~R. {Bareford} {et~al.}(2016){Bareford}, {Gordovskyy}, {Browning}, \& {Hood}}]{bareford_2016}
{Bareford}, M.~R., {Gordovskyy}, M., {Browning}, P.~K., \& {Hood}, A.~W. 2016, \bibinfo{title}{{Energy Release in Driven Twisted Coronal Loops},} \solphys, 291, 187, \dodoi{10.1007/s11207-015-0824-7}

\bibitem[{M.~R. {Bareford} \& A.~W. {Hood}(2015){Bareford} \& {Hood}}]{bareford_2015}
{Bareford}, M.~R., \& {Hood}, A.~W. 2015, \bibinfo{title}{{Shock heating in numerical simulations of kink-unstable coronal loops},} Philosophical Transactions of the Royal Society of London Series A, 373, 20140266, \dodoi{10.1098/rsta.2014.0266}

\bibitem[{M.~R. {Bareford} {et~al.}(2013){Bareford}, {Hood}, \& {Browning}}]{bareford_2013}
{Bareford}, M.~R., {Hood}, A.~W., \& {Browning}, P.~K. 2013, \bibinfo{title}{{Coronal heating by the partial relaxation of twisted loops},} \aap, 550, A40, \dodoi{10.1051/0004-6361/201219725}

\bibitem[{A. {Bhattacharjee} {et~al.}(2009){Bhattacharjee}, {Huang}, {Yang}, \& {Rogers}}]{bhattacharjee_2009}
{Bhattacharjee}, A., {Huang}, Y.-M., {Yang}, H., \& {Rogers}, B. 2009, \bibinfo{title}{{Fast reconnection in high-Lundquist-number plasmas due to the plasmoid Instability},} Physics of Plasmas, 16, 112102, \dodoi{10.1063/1.3264103}

\bibitem[{S. {Bingert} \& H. {Peter}(2011){Bingert} \& {Peter}}]{bingert_2011}
{Bingert}, S., \& {Peter}, H. 2011, \bibinfo{title}{{Intermittent heating in the solar corona employing a 3D MHD model},} \aap, 530, A112, \dodoi{10.1051/0004-6361/201016019}

\bibitem[{E.~J. Caramana {et~al.}(1998)Caramana, Shashkov, \& Whalen}]{caramana_1998}
Caramana, E.~J., Shashkov, M.~J., \& Whalen, P.~P. 1998, \bibinfo{title}{Formulations of {Artificial} {Viscosity} for {Multi}-dimensional {Shock} {Wave} {Computations},} Journal of Computational Physics, 144, 70, \dodoi{10.1006/jcph.1998.5989}

\bibitem[{P.~J. {Cargill} \& G.~W. {Pneuman}(1984){Cargill} \& {Pneuman}}]{cargill_1984}
{Cargill}, P.~J., \& {Pneuman}, G.~W. 1984, \bibinfo{title}{{Diamagnetic propulsion and energy balance of magnetic elements in the solar chromosphere and transition region},} \apj, 276, 369, \dodoi{10.1086/161621}

\bibitem[{F. {Chen} {et~al.}(2022){Chen}, {Rempel}, \& {Fan}}]{Chen_2022}
{Chen}, F., {Rempel}, M., \& {Fan}, Y. 2022, \bibinfo{title}{{A Comprehensive Radiative Magnetohydrodynamics Simulation of Active Region Scale Flux Emergence from the Convection Zone to the Corona},} \apj, 937, 91, \dodoi{10.3847/1538-4357/ac8f95}

\bibitem[{Y. {Chen} {et~al.}(2026){Chen}, {Peter}, {Przybylski}, {Chitta}, \& {Mandal}}]{chen_2025}
{Chen}, Y., {Peter}, H., {Przybylski}, D., {Chitta}, L.~P., \& {Mandal}, S. 2026, \bibinfo{title}{{A self-consistent 3D MHD model producing a solar blowout jet},} \aap, 706, A141, \dodoi{10.1051/0004-6361/202556660}

\bibitem[{X. Cheng {et~al.}(2023)Cheng, Priest, Li, Chen, Aulanier, Chitta, Wang, Peter, Zhu, Xing, Ding, Solanki, Berghmans, Teriaca, Aznar~Cuadrado, Zhukov, Guo, Long, Harra, Smith, Rodriguez, Verbeeck, Barczynski, \& Parenti}]{cheng_2023}
Cheng, X., Priest, E.~R., Li, H.~T., {et~al.} 2023, \bibinfo{title}{Ultra-high-resolution observations of persistent null-point reconnection in the solar corona,} Nature Communications, 14, 2107, \dodoi{10.1038/s41467-023-37888-w}

\bibitem[{G. Chouliaras \& V. Archontis(2025)Chouliaras \& Archontis}]{chouliaras_2025}
Chouliaras, G., \& Archontis, V. 2025, \bibinfo{title}{Magnetic {Flux} {Emergence} and {Solar} {Eruptions} in {Partially} {Ionized} {Plasmas},} \apj, 979, 53, \dodoi{10.3847/1538-4357/ada0af}

\bibitem[{G. Chouliaras {et~al.}(2023)Chouliaras, Syntelis, \& Archontis}]{chouliaras_2023}
Chouliaras, G., Syntelis, P., \& Archontis, V. 2023, \bibinfo{title}{Effects of Partial Ionization on Magnetic Flux Emergence in the Sun,} \apj, 952, 21, \dodoi{10.3847/1538-4357/acd9b2}

\bibitem[{J.~M. {da Silva Santos} {et~al.}(2022){da Silva Santos}, {Danilovic}, {Leenaarts}, {de la Cruz Rodr{\'\i}guez}, {Zhu}, {White}, {Vissers}, \& {Rempel}}]{Santos_2022}
{da Silva Santos}, J.~M., {Danilovic}, S., {Leenaarts}, J., {et~al.} 2022, \bibinfo{title}{{Heating of the solar chromosphere through current dissipation},} \aap, 661, A59, \dodoi{10.1051/0004-6361/202243191}

\bibitem[{Y. {Fan}(2001){Fan}}]{fan_2001}
{Fan}, Y. 2001, \bibinfo{title}{{The Emergence of a Twisted {\ensuremath{\Omega}}-Tube into the Solar Atmosphere},} \apjl, 554, L111, \dodoi{10.1086/320935}

\bibitem[{A.~J. {Finley} {et~al.}(2022){Finley}, {Brun}, {Carlsson}, {Szydlarski}, {Hansteen}, \& {Shoda}}]{finley_2022}
{Finley}, A.~J., {Brun}, A.~S., {Carlsson}, M., {et~al.} 2022, \bibinfo{title}{{Stirring the base of the solar wind: On heat transfer and vortex formation},} \aap, 665, A118, \dodoi{10.1051/0004-6361/202243947}

\bibitem[{H.~P. {Furth} {et~al.}(1963){Furth}, {Killeen}, \& {Rosenbluth}}]{furth_1963}
{Furth}, H.~P., {Killeen}, J., \& {Rosenbluth}, M.~N. 1963, \bibinfo{title}{{Finite-Resistivity Instabilities of a Sheet Pinch},} Physics of Fluids, 6, 459, \dodoi{10.1063/1.1706761}

\bibitem[{K. Galsgaard {et~al.}(2007)Galsgaard, Archontis, Moreno-Insertis, \& Hood}]{galsgaard_2007}
Galsgaard, K., Archontis, V., Moreno-Insertis, F., \& Hood, A.~W. 2007, \bibinfo{title}{The Effect of the Relative Orientation between the Coronal Field and New Emerging Flux. I. Global Properties,} \apj, 666, 516, \dodoi{10.1086/519756}

\bibitem[{C. {Gontikakis} {et~al.}(2008){Gontikakis}, {Contopoulos}, \& {Dara}}]{gontikakis_2008}
{Gontikakis}, C., {Contopoulos}, I., \& {Dara}, H.~C. 2008, \bibinfo{title}{{Distribution of coronal heating in a solar active region},} \aap, 489, 441, \dodoi{10.1051/0004-6361:20079086}

\bibitem[{B.~V. {Gudiksen} \& {\r{A}}. {Nordlund}(2005){Gudiksen} \& {Nordlund}}]{gudiksen_2005}
{Gudiksen}, B.~V., \& {Nordlund}, {\r{A}}. 2005, \bibinfo{title}{{An Ab Initio Approach to the Solar Coronal Heating Problem},} \apj, 618, 1020, \dodoi{10.1086/426063}

\bibitem[{F. {Guo} {et~al.}(2021){Guo}, {Giacalone}, \& {Zhao}}]{Fan_2021}
{Guo}, F., {Giacalone}, J., \& {Zhao}, L. 2021, \bibinfo{title}{{Shock Propagation and Associated Particle Acceleration in the Presence of Ambient Solar-Wind Turbulence},} Frontiers in Astronomy and Space Sciences, 8, 27, \dodoi{10.3389/fspas.2021.644354}

\bibitem[{A.~W. {Hood} {et~al.}(2012){Hood}, {Archontis}, \& {MacTaggart}}]{hood_2012}
{Hood}, A.~W., {Archontis}, V., \& {MacTaggart}, D. 2012, \bibinfo{title}{{3D MHD Flux Emergence Experiments: Idealised Models and Coronal Interactions},} \solphys, 278, 3, \dodoi{10.1007/s11207-011-9745-2}

\bibitem[{A.~W. {Hood} \& D.~W. {Hughes}(2011){Hood} \& {Hughes}}]{Hood_2011}
{Hood}, A.~W., \& {Hughes}, D.~W. 2011, \bibinfo{title}{{Solar magnetic fields},} Physics of the Earth and Planetary Interiors, 187, 78, \dodoi{10.1016/j.pepi.2011.04.010}

\bibitem[{Z. {Hou} {et~al.}(2022){Hou}, {Tian}, {Wang}, {Zhang}, {Song}, {Zheng}, {Chen}, {Chen}, {Bai}, {Chen}, {He}, {Song}, {Zhang}, {Hu}, {Dun}, {Zong}, {Song}, {Xu}, \& {Tan}}]{Hou_2022}
{Hou}, Z., {Tian}, H., {Wang}, J.-S., {et~al.} 2022, \bibinfo{title}{{Three-dimensional Propagation of the Global Extreme-ultraviolet Wave Associated with a Solar Eruption on 2021 October 28},} \apj, 928, 98, \dodoi{10.3847/1538-4357/ac590d}

\bibitem[{M. {Jarry} {et~al.}(2024){Jarry}, {Dresing}, {Rouillard}, {Plotnikov}, {Vainio}, {Palmroos}, {Kouloumvakos}, \& {Vuorinen}}]{Manon_2024}
{Jarry}, M., {Dresing}, N., {Rouillard}, A.~P., {et~al.} 2024, \bibinfo{title}{{The evolution of coronal shock wave properties and their relation with solar energetic particles},} \aap, 692, A92, \dodoi{10.1051/0004-6361/202451074}

\bibitem[{C. {Kanella} \& B.~V. {Gudiksen}(2017){Kanella} \& {Gudiksen}}]{kanella_2017}
{Kanella}, C., \& {Gudiksen}, B.~V. 2017, \bibinfo{title}{{Identification of coronal heating events in 3D simulations},} \aap, 603, A83, \dodoi{10.1051/0004-6361/201730748}

\bibitem[{A. {Kouloumvakos} {et~al.}(2023){Kouloumvakos}, {Vainio}, {Gieseler}, \& {Price}}]{Kouloumvakos_2023}
{Kouloumvakos}, A., {Vainio}, R., {Gieseler}, J., \& {Price}, D.~J. 2023, \bibinfo{title}{{The effect of shock wave properties on the release timings of solar energetic particles},} \aap, 669, A58, \dodoi{10.1051/0004-6361/202244363}

\bibitem[{A. {Kouloumvakos} {et~al.}(2022){Kouloumvakos}, {Kwon}, {Rodr{\'\i}guez-Garc{\'\i}a}, {Lario}, {Dresing}, {Kilpua}, {Vainio}, {T{\"o}r{\"o}k}, {Plotnikov}, {Rouillard}, {Downs}, {Linker}, {Malandraki}, {Pinto}, {Riley}, \& {Allen}}]{Kouloumvakos_2022}
{Kouloumvakos}, A., {Kwon}, R.~Y., {Rodr{\'\i}guez-Garc{\'\i}a}, L., {et~al.} 2022, \bibinfo{title}{{The first widespread solar energetic particle event of solar cycle 25 on 2020 November 29. Shock wave properties and the wide distribution of solar energetic particles},} \aap, 660, A84, \dodoi{10.1051/0004-6361/202142515}

\bibitem[{N. {Kumar} {et~al.}(2006){Kumar}, {Kumar}, \& {Singh}}]{Kumar_2006}
{Kumar}, N., {Kumar}, P., \& {Singh}, S. 2006, \bibinfo{title}{{Coronal heating by MHD waves},} \aap, 453, 1067, \dodoi{10.1051/0004-6361:20054141}

\bibitem[{P. {Kumar} {et~al.}(2023){Kumar}, {Karpen}, {Antiochos}, {DeVore}, {Wyper}, \& {Cho}}]{kumar_2023}
{Kumar}, P., {Karpen}, J.~T., {Antiochos}, S.~K., {et~al.} 2023, \bibinfo{title}{{Plasmoids, Flows, and Jets during Magnetic Reconnection in a Failed Solar Eruption},} \apj, 943, 156, \dodoi{10.3847/1538-4357/acaea4}

\bibitem[{V.~F. Kuropatenko(1967)Kuropatenko}]{kuropatenko_1967}
Kuropatenko, V.~F. 1967, \bibinfo{title}{---,} in Difference Methods for Solutions of Problems of Mathematical Physics, Vol.~1, ed. N.~N. Janenko (Providence, RI: American Mathematical Society), 116

\bibitem[{Y.~E. {Litvinenko}(2005){Litvinenko}}]{Litvinenko_2005}
{Litvinenko}, Y.~E. 2005, \bibinfo{title}{{Viscous Energy Dissipation by Flux Pile-Up Merging in the Solar Corona},} \solphys, 229, 203, \dodoi{10.1007/s11207-005-6882-5}

\bibitem[{N.~F. {Loureiro} {et~al.}(2007){Loureiro}, {Schekochihin}, \& {Cowley}}]{loureiro_2007}
{Loureiro}, N.~F., {Schekochihin}, A.~A., \& {Cowley}, S.~C. 2007, \bibinfo{title}{{Instability of current sheets and formation of plasmoid chains},} Physics of Plasmas, 14, 100703, \dodoi{10.1063/1.2783986}

\bibitem[{T. {Magara}(2004){Magara}}]{magara_2004}
{Magara}, T. 2004, \bibinfo{title}{{A Model for Dynamic Evolution of Emerging Magnetic Fields in the Sun},} \apj, 605, 480, \dodoi{10.1086/382148}

\bibitem[{S.~M. {Mahajan} {et~al.}(2001){Mahajan}, {Miklaszewski}, {Nikol'Skaya}, \& {Shatashvili}}]{Mahayan_2001}
{Mahajan}, S.~M., {Miklaszewski}, R., {Nikol'Skaya}, K.~I., \& {Shatashvili}, N.~L. 2001, \bibinfo{title}{{Formation and primary heating of the solar corona: Theory and simulation},} Physics of Plasmas, 8, 1340, \dodoi{10.1063/1.1350670}

\bibitem[{T. {Matsumoto}(2016){Matsumoto}}]{Matsumoto_2016}
{Matsumoto}, T. 2016, \bibinfo{title}{{Competition between shock and turbulent heating in coronal loop system},} \mnras, 463, 502, \dodoi{10.1093/mnras/stw2032}

\bibitem[{A. {Mignone} {et~al.}(2007){Mignone}, {Bodo}, {Massaglia}, {Matsakos}, {Tesileanu}, {Zanni}, \& {Ferrari}}]{Mignone_2007}
{Mignone}, A., {Bodo}, G., {Massaglia}, S., {et~al.} 2007, \bibinfo{title}{{PLUTO: A Numerical Code for Computational Astrophysics},} \apjs, 170, 228, \dodoi{10.1086/513316}

\bibitem[{T. {Miyagoshi} \& T. {Yokoyama}(2004){Miyagoshi} \& {Yokoyama}}]{Miyagoshi_2004}
{Miyagoshi}, T., \& {Yokoyama}, T. 2004, \bibinfo{title}{{Magnetohydrodynamic Simulation of Solar Coronal Chromospheric Evaporation Jets Caused by Magnetic Reconnection Associated with Magnetic Flux Emergence},} \apj, 614, 1042, \dodoi{10.1086/423731}

\bibitem[{R.~L. {Moore} {et~al.}(2010){Moore}, {Cirtain}, {Sterling}, \& {Falconer}}]{moore_2010}
{Moore}, R.~L., {Cirtain}, J.~W., {Sterling}, A.~C., \& {Falconer}, D.~A. 2010, \bibinfo{title}{{Dichotomy of Solar Coronal Jets: Standard Jets and Blowout Jets},} \apj, 720, 757, \dodoi{10.1088/0004-637X/720/1/757}

\bibitem[{K. Moraitis {et~al.}(2024)Moraitis, Archontis, \& Chouliaras}]{moraitis_2024}
Moraitis, K., Archontis, V., \& Chouliaras, G. 2024, \bibinfo{title}{Energy and helicity evolution in a flux emergence simulation,} A\&A, 690, A181, \dodoi{10.1051/0004-6361/202450924}

\bibitem[{F. {Moreno-Insertis} \& K. {Galsgaard}(2013){Moreno-Insertis} \& {Galsgaard}}]{moreno-insertis_2013}
{Moreno-Insertis}, F., \& {Galsgaard}, K. 2013, \bibinfo{title}{{Plasma Jets and Eruptions in Solar Coronal Holes: A Three-dimensional Flux Emergence Experiment},} \apj, 771, 20, \dodoi{10.1088/0004-637X/771/1/20}

\bibitem[{F. {Moreno-Insertis} {et~al.}(2008){Moreno-Insertis}, {Galsgaard}, \& {Ugarte-Urra}}]{moreno-insertis_2008}
{Moreno-Insertis}, F., {Galsgaard}, K., \& {Ugarte-Urra}, I. 2008, \bibinfo{title}{{Jets in Coronal Holes: Hinode Observations and Three-dimensional Computer Modeling},} \apjl, 673, L211, \dodoi{10.1086/527560}

\bibitem[{D. {N{\'o}brega-Siverio} {et~al.}(2017){N{\'o}brega-Siverio}, {Mart{\'\i}nez-Sykora}, {Moreno-Insertis}, \& {Rouppe van der Voort}}]{nobrega_2017}
{N{\'o}brega-Siverio}, D., {Mart{\'\i}nez-Sykora}, J., {Moreno-Insertis}, F., \& {Rouppe van der Voort}, L. 2017, \bibinfo{title}{{Surges and Si IV Bursts in the Solar Atmosphere: Understanding IRIS and SST Observations through RMHD Experiments},} \apj, 850, 153, \dodoi{10.3847/1538-4357/aa956c}

\bibitem[{D. {N{\'o}brega-Siverio} {et~al.}(2023){N{\'o}brega-Siverio}, {Moreno-Insertis}, {Galsgaard}, {Krikova}, {Rouppe van der Voort}, {Joshi}, \& {Madjarska}}]{Daniel_2023}
{N{\'o}brega-Siverio}, D., {Moreno-Insertis}, F., {Galsgaard}, K., {et~al.} 2023, \bibinfo{title}{{Deciphering Solar Coronal Heating: Energizing Small-scale Loops through Surface Convection},} \apjl, 958, L38, \dodoi{10.3847/2041-8213/ad0df0}

\bibitem[{D. {N{\'o}brega-Siverio} {et~al.}(2016){N{\'o}brega-Siverio}, {Moreno-Insertis}, \& {Mart{\'\i}nez-Sykora}}]{daniel_2016}
{N{\'o}brega-Siverio}, D., {Moreno-Insertis}, F., \& {Mart{\'\i}nez-Sykora}, J. 2016, \bibinfo{title}{{The Cool Surge Following Flux Emergence in a Radiation-MHD Experiment},} \apj, 822, 18, \dodoi{10.3847/0004-637X/822/1/18}

\bibitem[{D. {N{\'o}brega-Siverio} {et~al.}(2018){N{\'o}brega-Siverio}, {Moreno-Insertis}, \& {Mart{\'\i}nez-Sykora}}]{nobrega_2018}
{N{\'o}brega-Siverio}, D., {Moreno-Insertis}, F., \& {Mart{\'\i}nez-Sykora}, J. 2018, \bibinfo{title}{{On the Importance of the Nonequilibrium Ionization of Si IV and O IV and the Line of Sight in Solar Surges},} \apj, 858, 8, \dodoi{10.3847/1538-4357/aab9b9}

\bibitem[{Q. {Noraz} {et~al.}(2026){Noraz}, {Carlsson}, \& {Aulanier}}]{noraz_2025}
{Noraz}, Q., {Carlsson}, M., \& {Aulanier}, G. 2026, \bibinfo{title}{{Chromosphere of the quiet sun -- I. Shock and current-sheet dynamics and heating},} \aap, 705, A86, \dodoi{10.1051/0004-6361/202556375}

\bibitem[{D. {Nóbrega-Siverio} {et~al.}(2025){Nóbrega-Siverio}, {Joshi}, {Sola-Viladesau}, {Berghmans}, \& {Lim}}]{daniel_2025}
{Nóbrega-Siverio}, D., {Joshi}, R., {Sola-Viladesau}, E., {Berghmans}, D., \& {Lim}, D. 2025, \bibinfo{title}{Thin coronal jets and plasmoid-mediated reconnection - Insights from Solar Orbiter observations and Bifrost simulations,} A\&A, 702, A188, \dodoi{10.1051/0004-6361/202555357}

\bibitem[{E.~N. {Parker}(1957){Parker}}]{parker_1957}
{Parker}, E.~N. 1957, \bibinfo{title}{{Sweet's Mechanism for Merging Magnetic Fields in Conducting Fluids},} \jgr, 62, 509, \dodoi{10.1029/JZ062i004p00509}

\bibitem[{E.~N. {Parker}(1988){Parker}}]{parker_1988}
{Parker}, E.~N. 1988, \bibinfo{title}{{Nanoflares and the Solar X-Ray Corona},} \apj, 330, 474, \dodoi{10.1086/166485}

\bibitem[{H.~E. {Petschek}(1964){Petschek}}]{petschek_1964}
{Petschek}, H.~E. 1964, \bibinfo{title}{{Magnetic Field Annihilation},} in NASA Special Publication, ed. W.~N. {Hess}, Vol.~50, 425

\bibitem[{G.~W. {Pneuman}(1983){Pneuman}}]{pneuman_1983}
{Pneuman}, G.~W. 1983, \bibinfo{title}{{Ejection of magnetic fields from the sun - Acceleration of a solar wind containing diamagnetic plasmoids},} \apj, 265, 468, \dodoi{10.1086/160691}

\bibitem[{E. {Priest}(2014){Priest}}]{priest_2014}
{Priest}, E. 2014, {Magnetohydrodynamics of the Sun}, \dodoi{10.1017/CBO9781139020732}

\bibitem[{E.~R. {Priest} {et~al.}(1998){Priest}, {Foley}, {Heyvaerts}, {Arber}, {Culhane}, \& {Acton}}]{priest_1998}
{Priest}, E.~R., {Foley}, C.~R., {Heyvaerts}, J., {et~al.} 1998, \bibinfo{title}{{Nature of the heating mechanism for the diffuse solar corona},} \nat, 393, 545, \dodoi{10.1038/31166}

\bibitem[{A.~F. {Rappazzo} {et~al.}(2008){Rappazzo}, {Velli}, {Einaudi}, \& {Dahlburg}}]{rappazzo_2008}
{Rappazzo}, A.~F., {Velli}, M., {Einaudi}, G., \& {Dahlburg}, R.~B. 2008, \bibinfo{title}{{Nonlinear Dynamics of the Parker Scenario for Coronal Heating},} \apj, 677, 1348, \dodoi{10.1086/528786}

\bibitem[{J. {Reid} {et~al.}(2020){Reid}, {Cargill}, {Hood}, {Parnell}, \& {Arber}}]{reid_2020}
{Reid}, J., {Cargill}, P.~J., {Hood}, A.~W., {Parnell}, C.~E., \& {Arber}, T.~D. 2020, \bibinfo{title}{{Coronal energy release by MHD avalanches: Heating mechanisms},} \aap, 633, A158, \dodoi{10.1051/0004-6361/201937051}

\bibitem[{M. Rempel(2017)Rempel}]{Rempel_2017}
Rempel, M. 2017, \bibinfo{title}{Extension of the muram radiative mhd code for coronal simulations,} \apj, 834, 10, \dodoi{10.3847/1538-4357/834/1/10}

\bibitem[{D. {Ryu} {et~al.}(1995){Ryu}, {Jones}, \& {Frank}}]{Ryu_1995}
{Ryu}, D., {Jones}, T.~W., \& {Frank}, A. 1995, \bibinfo{title}{{Numerical Magnetohydrodynamics in Astrophysics: Algorithm and Tests for Multidimensional Flow},} \apj, 452, 785, \dodoi{10.1086/176347}

\bibitem[{A. {Schl{\"u}ter}(1957){Schl{\"u}ter}}]{schlueter_1957}
{Schl{\"u}ter}, A. 1957, \bibinfo{title}{{Solar radio emission and the acceleration of magnetic-storm particles},} in IAU Symposium, Vol.~4, Radio astronomy, ed. H.~C. {van de Hulst}, 356

\bibitem[{K. {Shibata} \& T. {Magara}(2011){Shibata} \& {Magara}}]{shibata_2011}
{Shibata}, K., \& {Magara}, T. 2011, \bibinfo{title}{{Solar Flares: Magnetohydrodynamic Processes},} Living Reviews in Solar Physics, 8, 6, \dodoi{10.12942/lrsp-2011-6}

\bibitem[{K. {Shibata} {et~al.}(2023){Shibata}, {Takasao}, \& {Reeves}}]{kengo_2023}
{Shibata}, K., {Takasao}, S., \& {Reeves}, K.~K. 2023, \bibinfo{title}{{Numerical Study on Excitation of Turbulence and Oscillation in Above-the-loop-top Region of a Solar Flare},} \apj, 943, 106, \dodoi{10.3847/1538-4357/acaa9c}

\bibitem[{K. Shibata \& S. Tanuma(2001)Shibata \& Tanuma}]{shibata_2001}
Shibata, K., \& Tanuma, S. 2001, \bibinfo{title}{Plasmoid-induced-reconnection and fractal reconnection,} Earth, Planets and Space, 53, 473, \dodoi{10.1186/BF03353258}

\bibitem[{J.~M. {Stone} \& M. {Edelman}(1995){Stone} \& {Edelman}}]{stone_1995}
{Stone}, J.~M., \& {Edelman}, M. 1995, \bibinfo{title}{{The Corrugation Instability in Slow Magnetosonic Shock Waves},} \apj, 454, 182, \dodoi{10.1086/176476}

\bibitem[{P.~A. {Sweet}(1958){Sweet}}]{sweet_1958}
{Sweet}, P.~A. 1958, \bibinfo{title}{{The Neutral Point Theory of Solar Flares},} in IAU Symposium, Vol.~6, Electromagnetic Phenomena in Cosmical Physics, ed. B.~{Lehnert}, 123, \dodoi{10.1017/S0074180900237704}

\bibitem[{S. {Takasao} {et~al.}(2015){Takasao}, {Matsumoto}, {Nakamura}, \& {Shibata}}]{Takasao_2015}
{Takasao}, S., {Matsumoto}, T., {Nakamura}, N., \& {Shibata}, K. 2015, \bibinfo{title}{{Magnetohydrodynamic Shocks in and above Post-flare Loops: Two-dimensional Simulation and a Simplified Model},} \apj, 805, 135, \dodoi{10.1088/0004-637X/805/2/135}

\bibitem[{S. {Takasao} \& K. {Shibata}(2016){Takasao} \& {Shibata}}]{Takasao_2016}
{Takasao}, S., \& {Shibata}, K. 2016, \bibinfo{title}{{Above-the-loop-top Oscillation and Quasi-periodic Coronal Wave Generation in Solar Flares},} \apj, 823, 150, \dodoi{10.3847/0004-637X/823/2/150}

\bibitem[{I. {Thaler} \& J.~M. {Borrero}(2023){Thaler} \& {Borrero}}]{Thaler_2023}
{Thaler}, I., \& {Borrero}, J.~M. 2023, \bibinfo{title}{{Photospheric signatures of retraction and reconnection in realistic magnetohydrodynamic simulations},} \aap, 673, A163, \dodoi{10.1051/0004-6361/202245701}

\bibitem[{G. {T{\'o}th}(2000){T{\'o}th}}]{Toth_2000}
{T{\'o}th}, G. 2000, \bibinfo{title}{{The {\ensuremath{\nabla}}{\textperiodcentered} B=0 Constraint in Shock-Capturing Magnetohydrodynamics Codes},} Journal of Computational Physics, 161, 605, \dodoi{10.1006/jcph.2000.6519}

\bibitem[{B. {van Leer}(1997){van Leer}}]{vanLeer_1997}
{van Leer}, B. 1997, \bibinfo{title}{{Towards the Ultimate Conservative Difference Scheme},} Journal of Computational Physics, 135, 229, \dodoi{10.1006/jcph.1997.5704}

\bibitem[{K. {Vanninathan} {et~al.}(2015){Vanninathan}, {Veronig}, {Dissauer}, {Madjarska}, {Hannah}, \& {Kontar}}]{Vanninathan_2015}
{Vanninathan}, K., {Veronig}, A.~M., {Dissauer}, K., {et~al.} 2015, \bibinfo{title}{{Coronal Response to an EUV Wave from DEM Analysis},} \apj, 812, 173, \dodoi{10.1088/0004-637X/812/2/173}

\bibitem[{Y. {Wang} \& T. {Yokoyama}(2020){Wang} \& {Yokoyama}}]{wang_2020}
{Wang}, Y., \& {Yokoyama}, T. 2020, \bibinfo{title}{{Simulation of Alfv{\'e}n Wave Propagation in the Magnetic Chromosphere with Radiative Loss: Effects of Nonlinear Mode Coupling on Chromospheric Heating},} \apj, 891, 110, \dodoi{10.3847/1538-4357/ab70b2}

\bibitem[{J. {Warnecke} \& S. {Bingert}(2020){Warnecke} \& {Bingert}}]{Warnecke_2020}
{Warnecke}, J., \& {Bingert}, S. 2020, \bibinfo{title}{{Non-Fourier description of heat flux evolution in 3D MHD simulations of the solar corona},} Geophysical and Astrophysical Fluid Dynamics, 114, 261, \dodoi{10.1080/03091929.2019.1670173}

\bibitem[{M.~L. Wilkins(1980)Wilkins}]{wilkins_1980}
Wilkins, M.~L. 1980, \bibinfo{title}{Use of artificial viscosity in multidimensional fluid dynamic calculations,} Journal of Computational Physics, 36, 281, \dodoi{10.1016/0021-9991(80)90161-8}

\bibitem[{L. {Yang} {et~al.}(2013){Yang}, {He}, {Peter}, {Tu}, {Zhang}, {Feng}, \& {Zhang}}]{yang_2013}
{Yang}, L., {He}, J., {Peter}, H., {et~al.} 2013, \bibinfo{title}{{Numerical Simulations of Chromospheric Anemone Jets Associated with Moving Magnetic Features},} \apj, 777, 16, \dodoi{10.1088/0004-637X/777/1/16}

\bibitem[{T. {Yokoyama} \& K. {Shibata}(1996){Yokoyama} \& {Shibata}}]{yokoyama_1996}
{Yokoyama}, T., \& {Shibata}, K. 1996, \bibinfo{title}{{Numerical Simulation of Solar Coronal X-Ray Jets Based on the Magnetic Reconnection Model},} \pasj, 48, 353, \dodoi{10.1093/pasj/48.2.353}

\bibitem[{A. {Zafar} {et~al.}(2023){Zafar}, {Ni}, {Lin}, \& {Ziegler}}]{Zafar_2023}
{Zafar}, A., {Ni}, L., {Lin}, J., \& {Ziegler}, U. 2023, \bibinfo{title}{{High {\ensuremath{\beta}} magnetic reconnection at different altitudes in the cool low solar atmosphere},} Physics of Plasmas, 30, 072902, \dodoi{10.1063/5.0149339}

\bibitem[{N.~I. Zahariev \& T.~M. Mishonov(2011)Zahariev \& Mishonov}]{Mishonov_2011}
Zahariev, N.~I., \& Mishonov, T.~M. 2011, \bibinfo{title}{Heating of the Solar Corona by Alfvén Waves: Self-Induced Opacity,} AIP Conference Series, 1356, 123, \dodoi{10.1063/1.3598101}

\bibitem[{S. {Zenitani} \& T. {Miyoshi}(2011){Zenitani} \& {Miyoshi}}]{zenitani_2011}
{Zenitani}, S., \& {Miyoshi}, T. 2011, \bibinfo{title}{{Magnetohydrodynamic structure of a plasmoid in fast reconnection in low-beta plasmas},} Physics of Plasmas, 18, 022105, \dodoi{10.1063/1.3554655}

\bibitem[{J. {Zhuleku} {et~al.}(2025){Zhuleku}, {Archontis}, \& {Moraitis}}]{zhuleku_2025}
{Zhuleku}, J., {Archontis}, V., \& {Moraitis}, K. 2025, \bibinfo{title}{{Recurrent Eruptions from the Emergence of a Toroidal Flux Tube},} \apj, 986, 47, \dodoi{10.3847/1538-4357/adda41}

\bibitem[{J. {Zhuleku} {et~al.}(2021){Zhuleku}, {Warnecke}, \& {Peter}}]{zhuleku_2021}
{Zhuleku}, J., {Warnecke}, J., \& {Peter}, H. 2021, \bibinfo{title}{{Stellar X-rays and magnetic activity in 3D MHD coronal models},} \aap, 652, A32, \dodoi{10.1051/0004-6361/202040192}

\end{thebibliography}
\bibliographystyle{aasjournalv7}


\end{document}